\documentclass[lettersize,journal]{IEEEtran}
\usepackage{amsmath,amsfonts}
\usepackage{algorithmic}
\usepackage{algorithm}
\usepackage{array}
\usepackage[caption=false,font=normalsize,labelfont=sf,textfont=sf]{subfig}
\usepackage{textcomp}
\usepackage{stfloats}
\usepackage{url}
\usepackage{verbatim}
\usepackage{graphicx}
\usepackage{cite}
\usepackage{multirow}
\usepackage{makecell}
\usepackage{threeparttable}
\usepackage{url}
\usepackage{bbding}
\usepackage{enumitem}
\usepackage[colorlinks, linkcolor=red, anchorcolor=red, citecolor=green]{hyperref}
\usepackage{multirow}
\usepackage{cite}
\usepackage{tabularx}
\usepackage{lipsum}
\usepackage{gensymb}
\usepackage{booktabs}
\usepackage{threeparttable}
\usepackage{caption}
\usepackage{subcaption}

\usepackage[table,xcdraw]{xcolor}
\usepackage{booktabs}
\usepackage{graphicx} % 引入相关图片处理功能的宏包

\begin{document}
\title{IPDnet: A Universal Direct-Path IPD Estimation Network for Sound Source Localization }

\author{Yabo~Wang, Bing~Yang, and Xiaofei~Li
        % <-this % stops a space
%\thanks{This work was supported by the Postdoctoral Science Foundation of China under Grant 2022M722848, and the Zhejiang Provincial Natural Science Foundation of China under Grant 2022XHSJJ008.}% <-this % stops a space
\thanks{Yabo Wang is with Zhejiang University and also with Westlake University, Hangzhou, China, e-mail: wangyabo@westlake.edu.cn.
Bing Yang and Xiaofei Li are with the School of Engineering, Westlake
University and also with the Institute of Advanced Technology, Westlake Institute for Advanced Study, Hangzhou, China
(e-mail: yangbing@westlake.edu.cn; lixiaofei@westlake.edu.cn). Yabo Wang and Bing Yang: equal contribution; Xiaofei Li: corresponding author.}
}



\maketitle

\begin{abstract}
Extracting direct-path spatial feature is crucial for sound source localization in adverse acoustic environments. This paper proposes the IPDnet, a neural network that estimates direct-path inter-channel phase difference (DP-IPD) of sound sources from microphone array signals. The estimated DP-IPD can be easily translated to source location based on the known microphone array geometry. First, a full-band and narrow-band fusion network is proposed for DP-IPD estimation, in which alternating narrow-band and full-band layers are responsible for estimating the rough DP-IPD information in one frequency band and capturing the frequency correlations of DP-IPD, respectively. Second, a new multi-track DP-IPD learning target is proposed for the localization of flexible number of sound sources. Third, the IPDnet is extend to handling variable microphone arrays, once trained which is able to process arbitrary microphone arrays with different number of channels and array topology. Experiments of multiple-moving-speaker localization are conducted on both simulated and real-world data, which show that the proposed full-band and narrow-band fusion network and the proposed multi-track DP-IPD learning target together achieves excellent sound source localization performance. Moreover, the proposed variable-array model generalizes well to unseen microphone arrays. Code is available on our github page \footnote{https://github.com/Audio-WestlakeU/FN-SSL}.

% named IPDnet which can estimate multiple DP-IPDs of multiple sources on a track-wise.   Moreover, the IPDnet for microphone array generalization version is designed which can be directly applied to localize sources with unseen microphone array configurations.
% The proposed IPDnet achieves multi-source localization across different microphone array configurations, enabling adaptation to a wider range of complex scenarios for sound source localization. Experiments are conducted on simulated and real datasets, and the results show that 1) the proposed  IPDnet outperforms other advanced methods on both simulated and real-world data; 2) The proposed array generalized version model  can be used on any unseen microphone array configuration and demonstrates superior performance.
% In contrast, self-supervised representation learning builds a universal model to benefit various tasks and domains, making it possible to take full advantage of real-world unlabeled data and reducing the requirement of data labeling.
%makes data collection in real world much easier.
%In contrast to other representation learning methods that aim to remove variabilities, ours is designed to preserve information for a wide range of downstream tasks.
\end{abstract}


\section{Introduction}
\IEEEPARstart{S}{ound} source localization (SSL) aims to estimate the position of one or multiple sound sources from microphone array signals. SSL is widely used in video conferencing and human-computer interaction. The spatial cues of SSL can also be used to boost the performance of speech enhancement and source separation tasks \cite{Lee2016DNNBasedFE, Chazan2019MultiMicrophoneSS}. 

Traditional SSL methods typically rely on estimating spatial features that are associated with the direct-path signal propagation in order to establish a mapping between features and source locations. Commonly used spatial features include time delay, inter-channel phase/level difference (IPD/ILD) \cite{4967888, Zhang2010ATM}, and relative transfer function (RTF) \cite{Braun2015NarrowbandDE, Wang2018SemiSupervisedLW}. Actually, the aforementioned spatial features can be straightforwardly estimated under ideal acoustic conditions (noise-free and anechoic condition). While estimating reliable features from microphone signals becomes challenging in real-world scenarios where noise, reverberation, and the presence of multiple moving sources introduce complexity. 
% Specifically, microphone signals consist of the direct-path signal, ambient noise, and reverberation. 
Noise and speech overlapping introduce uncertain acoustic distortions to the microphone signals and reverberation causes an overlap-masking effect or coloration of the originally anechoic signal \cite{Jeub2010ModelBasedDP}. Meanwhile, source movement introduces time-varying characteristics of spatial cues. 
% and aliasing effects to the direct-path propagation, which means the direct-path impulse response dynamically change over time and mixed with the impulse response from other sources. It is an additional challenge in separating and identifying the individual localization features from the mixture. 
These all result a significant decline in localization performance. 

In recent years, deep learning-based SSL approaches have been extensively developed, and show performance superiority over conventional methods  \cite{grumiaux2022survey, He2021NeuralNA, Grumiaux2021SaladnetSM, yang2021learning,DiazGuerra2020RobustSS,DiazGuerra2022DirectionOA, Fu2022IterativeSS,Yin2022MIMODoAnetMI,Yang2022SRPDNNLD,Kowalk2022GeometryAwareDE,Wang2023FramewiseMS,Cho2023SRSRPSB,WYBIS23}, particularly in challenging environments. 
% Compared with conventional methods, deep learning-based approaches demonstrate superior adaptability to diverse acoustic scenarios encountered within the training data. 
The superiority stems from the capacity of neural networks to learn complex patterns and subtle differences in acoustic signals. This work proposes a new SSL network for multiple-moving-source localization in the presence of noise and reverberation, which is an extended version of our previous conference paper, i.e. FN-SSL \cite{WYBIS23}. FN-SSL is a full-band and narrow-band fusion network proposed for single-source localization with two microphones, which achieved excellent SSL performance due to its efficient network architecture. In this work, we extend FN-SSL for multi-microphone and multi-source localization, and propose a new learning target. Moreover, based on the extended network, a new variable-array model is proposed which can be applied to variable microphone arrays with different array topology. 

% Designing a universal neural network capable of adapting to complex acoustic conditions and being used across different microphone array configurations is essential for the field of SSL. The proposed method IPDnet is an extended version of our previous work \cite{WYBIS23}, which can estimate multiple direct-path IPDs (DP-IPDs). IPDnet is a universal SSL neural network that employs fusion of full-band and narrow-band information to estimate multi-microphone, multi-source DP-IPDs. Meanwhile, it can also be directly applied to localize sources with unseen microphone array configurations. 
% However, these methods rely heavily on data and necessitate the acquisition of a substantial volume of annotated, device-specific training data, incurring significant costs. Furthermore, once trained, most deep learning-based models are restricted to working with specific microphone arrays, imposing limitations on their applicability across different setups.

Specifically, the contributions of this work are as follows:
\subsubsection{Full-band and narrow-band fusion SSL network.} The proposed network is motivated by the recently proposed speech enhancement networks \cite{tesch2022insights, Yang2022McNetFM,quan2023spatialnet}, in which full-band layers and narrow-band layers are cascaded for predicting the clean speech signal, which shows a large performance superiority for speech enhancement, and is now becoming a new research trend. 
The proposed network takes as input the (short-time Fourier transform, STFT of) multichannel microphone signals, and predicts the direct-path IPD (DP-IPD) as localization feature.   
Compared to processing the noisy IPD \cite{Nguyen2021SALSALiteAF} or the noisy spatial spectrum \cite{DiazGuerra2020RobustSS,DiazGuerra2022DirectionOA}, the microphone signals preserve the natural properties of noise and reverberation, such as the spatial-diffuseness of noise and late reverberation, and it is more effective to leverage these properties to remove them. In the proposed network, full-band/narrow-band layers process the time frames/frequencies independently, and all the time
frames/frequencies share the same network weights. In narrow-band, there are rich information for extracting localization features, which are largely leveraged in conventional methods. For example, localization features are extracted by narrow-band channel identification in \cite{7533416}, by coherence test in \cite{Mohan2008LocalizationOM}, and by direct-path dominance test in \cite{Nadiri2014LocalizationOM}. The narrow-band layers processes the along-time sequences to focus on learning these narrow-band information.
The full-band layers processes the along-frequency sequence to focus on learning the full-band correlation of spatial cues, such as the linear relation of DP-IPD to frequency.
% Full-band and narrow-band fusion to estimate the spatial feature is motivated by the recently proposed speech enhancement networks , and has been demonstrated to be effective in our previous work. 
% The alternating full-band and narrow-band layers are tasked with learning the full-band correlation and narrow-band extraction of DP-IPD. Estimating DP-IPDs in a track-wise manner proves effective in distinguishing DP-IPDs from different sources. Experimental results demonstrate that the proposed network is capable of training with any number of microphone array data and notably outperforms other advanced methods on both simulated and real-world data.
\subsubsection{Mutli-track DP-IPD learning target.} 

Popular SSL learning targets include location classification \cite{Grumiaux2021SaladnetSM}, location regression \cite{Shimada2021MultiACCDOALA} and spatial spectrum regression \cite{He2021NeuralNA}. In this work, we propose to use multi-track DP-IPD as the learning target for multi-source localization. DP-IPD stands for the IPD of the direct-path propagation, which is theoretically related to the microphone array geometry and source location, and thus it can be easily translated to source location with known array geometry. In this work, we estimate the source location by simply matching the estimated DP-IPD with theoretical DP-IPD of candidate locations. Multi-track DP-IPD means we let the network output/estimate the DP-IPD of multiple sources simultaneously. DP-IPD is a signal-level localization feature, which can be estimated from microphone signals based on only signal-level information. By contrast, other targets are array-dependent, and require one further step of conversion from localization feature to target, which may arise more difficulty. This is verified by our experiments conducted in Section \ref{sec:target}.
% Our experiments show that, in the present SSL framework that takes as input the microphone signals, the proposed DP-IPD target achieves noticeably better performance than other targets mentioned above. 

DP-IPD is well defined for active sound source, but not for non-source frames. One straightforward representation for non-source is an all-zero vector. However, learning the output space combined by DP-IPDs and the all-zero vector may be not easy. This work proposes taking the DP-IPD mean point of the whole localization space as the representation for non-source, with which it is more easier to switch between source and non-source frames, as shown in Fig.~\ref{fig:newtarget}. 

% how to use DP-IPD as the learning target for microphone arrays with more than two microphones remains an ambiguous issue. In this paper, a solution is proposed that
% selecting a reference channel from the multi-channel microphone signals, with the network's target designed as the DP-IPDs between the remaining channels and the reference channel. Additionally, A track-wise full-band and narrow-band fusion network named IPDnet is proposed to estimate the multiple DP-IPDs of multiple sources.
% In the task of estimating DP-IPD, a straightforward target setting is to use all-zero vector to represent silent frames. Although the all-zero vector and DP-IPD can be easily distinguished, there is no inherent correlation between all-zero vector and the DP-IPD vector. For neural networks, it is more challenging to switch quickly between all-zero vector and the DP-IPD vector compared to switching between different DP-IPD vectors. In this paper, the spatial coherence of the diffuse signal is employed as the target for silent frames. It can be considered an integration across various candidate directions of the DP-IPD in spatial terms. Compared to all-zero vector, the proposed target for silent frames has an intrinsic relationship with DP-IPD, making it easily distinguishable. Experimental results demonstrate that averaged DP-IPD vector of all candidate directions  leads to superior localization performance compared to employing the all-zero vector.
\subsubsection{Variable Array SSL} Most of existing SSL networks are array-dependent, namely training and test using the same array. 
% cannot generalize to unseen microphone array configurations, 
For a new array, the network needs to be retrained, which is time-consuming. Especially, when using real data for training, collecting a large amount of annotated data for a new array is even much more time consuming. In this paper, a variable-array SSL model is proposed, once trained which can be directly used for any unseen microphone array. Specifically, microphones are processed pair-wisely, and the mean pooling of pair-wise hidden units is used for the communication between microphone pairs. This 
pair-wise processing plus mean pooling scheme can handle variable number of microphones, and it is motivated by the variable-array speech enhancement networks \cite{luo2020end,Taherian2021OneMT,Zhang2021MicrophoneAG,Yoshioka2021VarArrayAC}. The network outputs the DP-IPD estimation for varying number of microphone pairs. As DP-IPD estimation is a signal-level task, one network can easily handle the task for different arrays. At the test stage, the estimated DP-IPDs can be used for source localization with the known array geometry. 
% Although splitting the microphone array into microphone pairs and learning the DP-IPD of each pair dependently can also be used for microphone array generalization, it lacks the interaction of spatial information between microphone pairs. A simple yet effective operation is designed to increase interaction between channels, which is concatenating the averaged embedding of all microphone pairs to each  microphone pair.  The experiment shows that the proposed method achieves superior generalization  on various type of microphone arrays.

Experiments have been conducted on both simulated and real dataset, which demonstrate that the proposed learning target outperforms all comparison targets, and the proposed method as a whole outperforms recently proposed baseline methods by a large margin. Moreover, the proposed variable-array model can generalize well to unseen simulated and real microphone arrays. 

The rest of this paper is organized as follows. Section \ref{sec:related} presents an overview of related works in the literature. Section \ref{sec:Formulation} defines the problem of multiple moving source localization. Section \ref{sec:method} details the proposed method. Section \ref{sec:exp} gives the experimental results and discussions. Finally, Section 
\ref{conclusion} concludes the paper and suggesting directions for future research.
% \usepackage{multirow}
\begin{table*}[]
\renewcommand\arraystretch{1.2}
\tabcolsep0.05in
\centering
\caption{Brief overview of deep-learning-based sound source localization methods.}
\label{tab:related}
\begin{tabular}{ccccccc}
\toprule \toprule
\multirow{2}{*}{\textbf{Method}} & \multirow{2}{*}{\textbf{Year}} & \multirow{2}{*}{\textbf{Input Feature}} & \multirow{2}{*}{\textbf{Target}}        & \textbf{Multi-source}       & \textbf{Frame-wise}       & \textbf{Variable-array}     \\
                                 &                                &                                         &                                         & \textbf{(vs. Single-source)} & \textbf{(vs. Chunk-wise)} & \textbf{(vs. Fixed-array )} \\ \midrule
                             \cite{He2021NeuralNA}     & 2021                           & Mag + Phase                             & Spatial spectrum regression             & \Checkmark                            & \XSolid                         & \XSolid                           \\
                             \cite{Grumiaux2021SaladnetSM}     & 2021                           & Intensity vector                        & Multi-class location classification     & \Checkmark                            & \Checkmark                         & \XSolid                           \\
                             \cite{yang2021learning}    & 2021                           & Mag + Phase                             & DP-RTF regression                       & \XSolid                            & \XSolid                         & \Checkmark(2-channel)                \\
                             \cite{DiazGuerra2020RobustSS}    & 2021                           & SRP-PHAT Spectrogram                    & Location regression                     & \XSolid                            & \Checkmark                         & \XSolid                           \\
                            \cite{DiazGuerra2022DirectionOA}     & 2022                           & SRP-PHAT Spectrogram                    & Location regression                     & \XSolid                            & \Checkmark                         & \XSolid                           \\
                            \cite{Fu2022IterativeSS}     & 2022                           & STFT Coefficients                       & Spatial spectrum regression             & \Checkmark                            & \XSolid                         & \XSolid                           \\
                              \cite{Yin2022MIMODoAnetMI}   & 2022                           & Mag + IPD                               & Multi-track spatial spectrum regression & \Checkmark                            & \XSolid                         & \XSolid                           \\
                              \cite{Yang2022SRPDNNLD}   & 2022                           & Mag + Phase                             & Mixed DP-IPD regression                 & \Checkmark                            & \Checkmark                         & \XSolid                           \\
                               \cite{Kowalk2022GeometryAwareDE}  & 2022                           & GCC-PHAT + Array Geometry               & Location classification                 & \XSolid                            & \XSolid                         & \Checkmark(constant-channel)         \\
                              \cite{Wang2023FramewiseMS}   & 2023                           & MFCC and Mel features                   & Multi-class location classification     & \Checkmark                            & \Checkmark                         & \XSolid                            \\ 
                             \cite{Cho2023SRSRPSB}    & 2023                           & SRP-PHAT Spectrogram                    & Location regression                     & \XSolid                            & \Checkmark                         & \XSolid                           \\
                               \cite{WYBIS23}  & 2023                           & STFT Coefficients                       & DP-IPD regression                       & \XSolid                            & \Checkmark                         & \XSolid                           \\ \midrule
Proposed                         & -                              & STFT Coefficients                       & Multi-track DP-IPD regression           & \Checkmark                            & \Checkmark                         & \Checkmark   
\\ \bottomrule \bottomrule
\end{tabular}
\end{table*}

\section{Related Works}\label{sec:related}
\subsection{Deep Learning Based Sound Source Localization}
In recent years, significant research progress has been made in the field of sound source localization using neural networks \cite{grumiaux2022survey, He2021NeuralNA, Grumiaux2021SaladnetSM, yang2021learning,DiazGuerra2020RobustSS,DiazGuerra2022DirectionOA, Fu2022IterativeSS,Yin2022MIMODoAnetMI,Yang2022SRPDNNLD,Kowalk2022GeometryAwareDE,Wang2023FramewiseMS,Cho2023SRSRPSB,WYBIS23}. Table \ref{tab:related} provides a chronological overview of some representative sound source localization methods. Although these methods normally achieve promising SSL performance, they can only handle certain limited tasks, in terms of the number of sources, frame-wise or block-wise SSL, fixed array or variable array. Where multi-source localization sometimes requires especially designed output format, frame-wise methods output SSL result for each time frame and are suitable for online and moving source localization, variable-array models can be applied to unseen microphone arrays.  
% only focus on one type of problem in sound source localization for research, meanwhile, the generalization ability  of them are closely related to the data in the training set.

Various network architectures have been adopted for SSL, among which convolutional neural networks (CNN) \cite{He2021NeuralNA,DiazGuerra2022DirectionOA,Cho2023SRSRPSB,Fu2022IterativeSS} and convolutional recurrent neural Networks\cite{yang2021learning, Yang2022SRPDNNLD,Wang2023FramewiseMS} (CRNN) 
% for SSL include fully connected (FC) neural network \cite{Kowalk2022GeometryAwareDE}, convolutional neural networks (CNN)\cite{He2021NeuralNA,DiazGuerra2022DirectionOA,Cho2023SRSRPSB,Fu2022IterativeSS}, convolutional recurrent neural Networks\cite{yang2021learning, Yang2022SRPDNNLD,Wang2023FramewiseMS} (CRNN) and recurrent neural network (RNN) \cite{Yin2022MIMODoAnetMI,WYBIS23,TDOA_LSTM_ICASSP19}. 
% Among these methods, CNN and CRNN 
are the most commonly used networks. These networks are all designed to process all the frequencies together. 
% Convolutional layers are utilized to extract local spatial information, while RNNs are employed to capture the long-term temporal context of this spatial information. 
The network input can be in the signal level, such as the time-domain signal \cite{Vecchiotti2019EndtoendBS}, the STFT coefficients \cite{Fu2022IterativeSS,WYBIS23} or the magnitude and phase of STFT coefficients \cite{He2021NeuralNA,yang2021learning,Yang2022SRPDNNLD}, or in the feature level, such as IPD, IID, the generalized cross-correlation (GCC) function \cite{Xiao2015ALA, Ma2017ExploitingDN, TDOA_LSTM_ICASSP19}  and noisy spatial spectrum \cite{DiazGuerra2020RobustSS,DiazGuerra2022DirectionOA,Cho2023SRSRPSB}. 
% Recent studys have shown that not performing explicit feature extraction on the signal allows the network to more effectively extract the most informative spatial features for SSL \cite{He2021NeuralNA, Vecchiotti2019EndtoendBS, WYBIS23}, that the signal contains the natural properties of noise and reverberation (such as the spatial-diffuseness). 

According to the learning target, SSL methods are classified as  feature/location regression or location classification methods. Feature/location regression methods estimate the localization feature (such as DP-RTF, DP-IPD and inter-channel time difference (ITD)) \cite{TDOA_LSTM_ICASSP19, yang2021learning ,Yang2022SRPDNNLD, WYBIS23, Yang2021EnhancingDR, Tang2019SupervisedCE,Pak2019SoundLB} or directly estimate source location \cite{He2021NeuralNA,DiazGuerra2020RobustSS,DiazGuerra2022DirectionOA,Fu2022IterativeSS,Cho2023SRSRPSB} from the noisy signal or noisy localization features. Most works output the feature/location for one source, and few works study how to extend feature/location regression to multiple sources. Location classification methods \cite{Grumiaux2021SaladnetSM,Wang2023FramewiseMS,Kowalk2022GeometryAwareDE,Yin2022MIMODoAnetMI,Adavanne2017DirectionOA,Nguyen2020RobustSC,Adavanne2018SoundEL} take candidate locations as classes, and multi-source localization can be easily conducted as a multi-class classification task \cite{Grumiaux2021SaladnetSM,Adavanne2017DirectionOA,Nguyen2020RobustSC,Adavanne2018SoundEL}. 
% Taking SSL task as a multi-class classification task can extend such methods to multiple sources localization \cite{Grumiaux2021SaladnetSM,Adavanne2017DirectionOA,Nguyen2020RobustSC,Adavanne2018SoundEL}. 
Due to the non-orthogonal relationship between adjacent locations, the classification output often exhibits a blurred response around the main peak which degrades the localization performance.

Based on Table \ref{tab:related} and the above overview, it is clear that the proposed method is totally different from existing works in both network architecture and learning target. 
% As mentioned, localization in real-world environments is complex. Despite numerous deep learning-based studies on SSL, it can be observed from Table \ref{tab:related} that most methods focus on specific sub-problems of SSL (such as moving source or multiple sources). Additionally, the data collection cost for deep learning is high, and models trained on specific array configurations cannot generalize well to unseen array configurations. This paper aims to propose a general neural network to address these issues of SSL. 
\subsection{Variable Array SSL}
% Most of existing SSL networks are array-dependent, namely training and test use the same array.
% Deep learning has achieved significant performance improvement in sound source localization tasks. Most deep learning based methods
% train the network for a dedicated microphone array and apply the network to the same array. While these trained model will not perform well on other unseen arrays due to the mapping from localization features to source positions is related to the array topology. 
Mapping from microphone signals and/or localization features to source location is intrinsically an array-dependent problem, which requires to know the geometry of microphone array or uses a fixed microphone array. 
% The neural network cannot learn the relationship between input features and the direction of arrival (DoA) without the information of microphone array topology. 
% Few works concern the generalization to unseen arrays. Ma \textit{et al.} \cite{Ma2017ExploitingDN} use the multi-condition training (MCT) to deal with the head mismatch between training and test. 
In \cite{Kowalk2022GeometryAwareDE}, by also taking as input the microphone array geometry along the localization feature to the network, the network can perform SSL for variable arrays. However, limited by the fixed input size, one network can only process variable arrays with the same number of microphones. In our previous works \cite{yang2021learning, Yang2022SRPDNNLD}, the clean localization feature, i.e. DP-RTF or DP-IPD, is taken as the network output, and 2-channel array is processed, for which case one network can be directly trained with variable arrays and test on unseen arrays. In this work, the proposed variable-array model can handle any microphone array without the limit of number of microphones. 

% a mechanism for information interaction between microphone pairs was designed. The information of all microphone pairs is provided to each microphone as a complement of spatial features to estimate the DP-IPD which can be flexible used on unseen microphone arrays.

% In our previous works \cite{yang2021learning, Yang2022SRPDNNLD,WYBIS23}, DP-RTF/DP-IPD is used as the target of the network. 
% Since DP-RTF/DP-IPD  is the feature between microphone pairs, the estimation of it is independent of the microphone array topology, and thence the DP-RTF/DP-IPD learning network can be directly applied to unseen arrays. By taking the inner product with the theoretical values of DP-RTF/DP-IPD corresponding to each azimuth, the mapping relationship between DP-RTF/DP-IPD and Direction of Arrival (DOA) can be established. Although using DP-RTF/DP-IPD can generalize to unseen microphone arrays , a pair of microphones provides fewer spatial information than a full multi-microphone array. Further investigation is required about how to use DNN to deal with the array generalization problem.
% Motivated by the array generalization methods in the speech enhancement task \cite{Taherian2021OneMT,Zhang2021MicrophoneAG,Yoshioka2021VarArrayAC}, in which averaging the features of microphone pairs can effectively utilize the spatial information among all microphone pairs. 

\section{Problem Formulation}\label{sec:Formulation}
Assuming there are multiple sound sources in a closed environment with noise and reverberation. 
% source at a direction $\theta$ is observed by the microphone array. 
The multichannel signals recorded by a microphone array are denoted as
\begin{equation}
x_m(t)=\sum_{k=1}^K a_m\left(t,\theta_k\right)*s_k(t)+v_m(t),
\end{equation}
where $m \in[1, M]$, $k \in[1, K]$ and $t \in[1, T]$ represent the indices of microphones, sound sources and time samples, respectively. 
% Note that in this task, $M$ is not a fixed value that we use $M$ to denote the number of microphones for presentation simplicity. 
% $x_m(t)$, $s_k(t)$ and represent the microphone signal, source and noise signals, respectively. 
As for the $k$-th source, $s_k(t)$, $\theta_k$, and $a_m\left(t, \theta_k\right)$ represent the source signal, the direction of arrival (DOA), and the direct-path response (within the room impulse response, RIR) to the $m$-th microphone, respectively, and * denotes convolution. The noise signal $v_m(t)$ includes both ambient noise and the reflections/reverberation of sources.

Applying the short-time Fourier transform (STFT), the multichannel signals are expressed as 
\begin{equation}
X_m(n, f)=\sum_{k=1}^K A_m\left(f, \theta_k\right) S_k(n, f)+V_m(n, f),
\end{equation}
where $n \in[1, N]$, $f \in[1, F]$ represent the time frame index and frequency index, respectively. Here $X_m(n, f)$, $S_k(n, f)$ and $V_m(n, f)$ are the STFT coefficients of microphone, source and noise signals, respectively. $A_m\left(f, \theta_k\right)$ is the transfer function (Fourier transform) of the direct-path response. 
% The room transfer function consists of the direct-path and the reflection-path component.
% \begin{equation}
% A_m(n,f, \theta)=A_m^{\mathrm{d}}(n,f, \theta)+A_m^{\mathrm{r}}(n,f, \theta),
% \end{equation}
% where $A_m^{\mathrm{d}}(t,k, \theta)$ and $A_m^{\mathrm{r}}(t,k, \theta)$ represent the direct-path and reflection-path parts, respectively. 
The direct-path relative transfer function (DP-RTF) of two microphones encodes the direct-path IPD and ILD within its phase and amplitude, respectively, and it is thus a reliable localization feature 
\begin{equation}
   B_m(f,\theta_k)={A_m(f, \theta_k)} / {A_r(f, \theta_k)},
\end{equation}
where $r$ is the index of one selected reference channel. For simplicity, only the DP-IPD, i.e. the phase part $\angle B_m(f,\theta_k)$, is employed and learned for SSL in this work.

In the free and far field, for one given microphone pair and source DOA $\theta$ (relative to the microphone pair), the complex-valued DP-IPD can be theoretically computed as 
\begin{equation}\label{dp-ipd}
 \tilde{B}(f,\theta) = e^{-j 2\pi v_f d \text{cos}(\theta)/c}
\end{equation}
where $v_f$ is the frequency in Hz, $d$ is the microphone distance, $c$ is sound speed in air, and $d\text{cos}(\theta)/c$ is the time difference of arrival (TDOA) from the direction of $\theta$ to the two microphones. 

\begin{figure}
    \centering
    \includegraphics[scale=0.55]{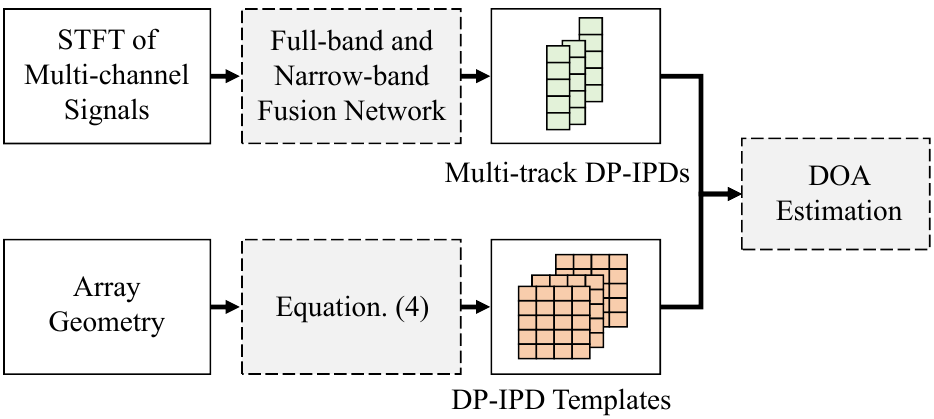}
    \caption{Block diagram of the proposed method.}
    \label{fig:template}
\end{figure}

In this work, sound source localization amounts to using a neural network to  estimate the DP-IPDs from the multichannel microphone signals.  As for $M$ microphones, one of microphone is selected as the reference channel, and the network predicts the DP-IPDs of the $M-1$ microphone pairs (other channels relative to the reference channel). Then, the DOA estimation can be obtained by simply matching the predicted DP-IPDs with the DP-IPD templates (namely the theoretical DP-IPDs of a set of pre-defined candidate directions), as shown in Fig. \ref{fig:template}. Specifically, the inner product between the predicted DP-IPD vector with the theoretical DP-IPD vector are computed, and the candidate direction with the maximum product value is taken as the DOA estimation. The training targets of DP-IPD and the DP-IPD templates are all computed using Eq.~(\ref{dp-ipd}). 

Moreover, the proposed DP-IPD estimation network is designed to handle more realistic and complex applications in the following aspects: 

\begin{itemize} [leftmargin=*] 
\item Flexible number of sound sources. We consider the number of source sources is flexible and time varying. However, we rely on a strong and reasonable assumption that at most $K$ (e.g. 2 in this work) sources present at one time, based on which a fixed number of $K$ tracks of DP-IPD are predicted. The DP-IPD estimations could belong to different sources at different times in one track. A trivial DP-IPD value is set for non-source. Overall, the network can detect and localize 0 to $K$ sources at one time. 
\item Moving sound source. For moving source, the DOA, i.e. $\theta_k$, and also its transfer function $A_m\left(f, \theta_k\right)$ are time-dependent/varying. To address this, we frame-wisely predict the DP-IPD and localize the sources, in either an online (causal) or an offline (non-causal) way.
\item Variable microphone arrays. The network can be designed to work for variable arrays with different topology and number of channels. One network is trained using many different arrays, and it predicts the DP-IPDs for the varying $M-1$ microphone pairs. Then, the network can be used for DP-IPD estimation with an arbitrary test array, for which the DP-IPD templates can be theoretically computed during test. This way disentangles the DP-IPD estimation step and the localization step (namely template matching), and thus forms an universal sound source localization network. This is reasonable considering the facts that the DP-IPD estimation step could be an array-independent signal-level task, and the localization step is array-dependent but theoretically simple. The accuracy of DP-IPD estimation is critical for localization, as perfect DP-IPD estimation leads to nearly perfect localization.
\end{itemize}
The number of microphones and sources, i.e. $M$ and $K$, and the DOA $\theta_k$ could all be varied either for different settings or along time, but we will not specify the variation in the following for notational simplicity, unless otherwise stated.  

% due to the DP-IPD is the feature of microphone pairs,  multi-microphone array (M$>$2) corresponds to multiple DP-RTFs. Once a reference channel is selected, the problem of this work can be defined as: using the $M$ channel microphone signals to estimate $m-1$ DP-RTFs. As IPD is more discriminative than ILD. Hence, this work only take the DP-IPD as the localization feature. The DP-IPD can be represented as
% $\angle B^{\mathrm{d}}(n,f, \theta)$. This work aims to estimate the multiple DP-IPDs of multiple sound sources through neural networks. Once the DP-IPDs are obtained, the mapping from DP-IPD to DOA (Direction of Arrival) is illustrated in Fig. \ref{fig:template}. The localization space is divided into multiple candidate positions according to a predetermined resolution. Then, utilizing the topology information of the microphone array, the DP-IPD for each candidate position is calculated to form a DP-IPD template. By conducting an inner product between the estimated DP-IPD and the template, the final DOA is determined.

\section{Method}\label{sec:method}
\begin{figure*}
    \centering
    \includegraphics[scale=0.72]{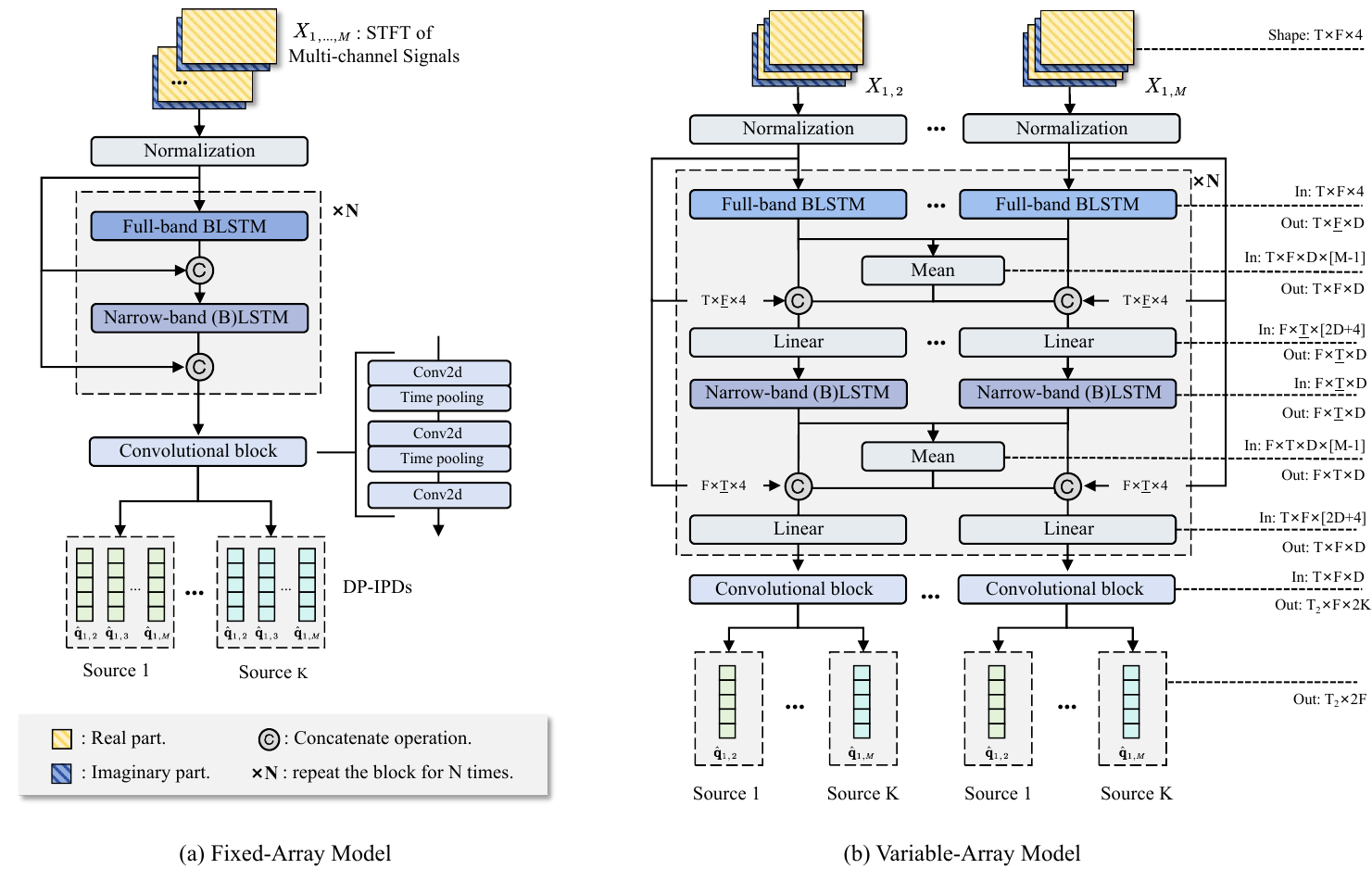}
    \caption{Network architecture for the proposed fix-array and variable-array models. 
The data organization is in the format of: number of sequences $\times$ \underline{sequence length} $\times$ feature dimension.}
    \label{fig:method}
\end{figure*}

This section presents the proposed IPDNet. Two versions are proposed for a fixed microphone array and variable arrays, whose network architectures are shown in Fig. \ref{fig:method} (a) and (b), respectively.  
% Fig. \ref{fig:method} (a) is the one-array-dedicated version which used to process the multichannel signals that the microphone array topology is determined. Fig. \ref{fig:method} (b) is the version of microphone array generalization that the trained model can be used on microphone arrays of any topology. The specific model structure is described as follows.

\subsection{Learning Target}
The proposed network takes as input the STFT of multichannel recordings, and outputs/predicts the DP-IPD features. To enable the optimization of real-value networks, following the setting in our prevision works \cite{yang2021learning,Yang2022SRPDNNLD,WYBIS23}, the learning target is set as the real ($\mathcal{R}$) and imaginary ($\mathcal{I}$) parts of complex-valued DP-IPD (concatenated along frequencies, for one microphone pair) as
\begin{equation}
\begin{aligned}
\mathbf{q}(\theta)= 
 [&\mathcal{R}\{\tilde{B}(1, \theta)\}, \ldots, \mathcal{R} \{\tilde{B}(F, \theta)\}, \\
& \mathcal{I} \{\tilde{B}(1, \theta)\}, \ldots, \mathcal{I} \{ \tilde{B}(F, \theta)\}]^{\top} \in \mathbb{R}^{2 F},
\end{aligned}
\end{equation}
where $^{\top}$ denotes vector transpose. The output activation layer is set as \emph{tanh} to predict DP-IPD. 
% At inference, the inner product between the predicted DP-IPD vector of one frame and the DP-IPD vector of candidate directions are computed, and the candidate direction with the largest inner product is taken as the localization result. The candidate directions are sampled in the whole localization space.

\begin{figure}
    \centering
    \includegraphics[scale=0.75]{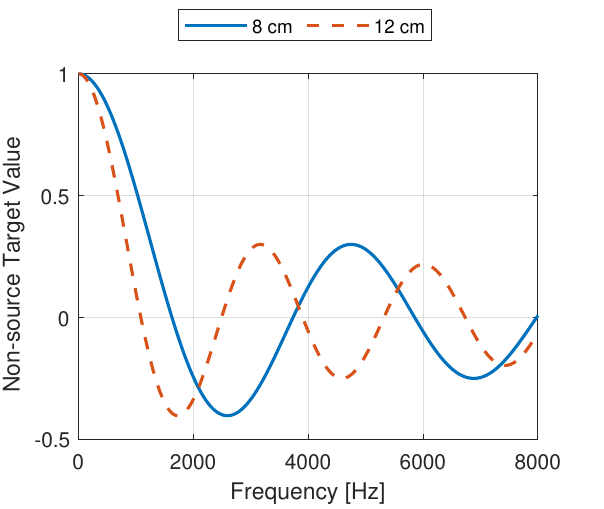}
    \caption{Examples of non-source target for two different microphone distances. }
    \label{fig:target}
\end{figure}

$\mathbf{q}(\theta)$ defines the DP-IPD target vector for one sound source presents at  DOA $\theta$, and across all DOAs it forms an DP-IPD manifold. However, it is not straightforward to define the target for non-source frames. In our previous works, an all-zero vector is used as the target for non-source frames, which however seems not meaningfully correlated to the DP-IPD vector. The network needs to learn the output space expanded by the DP-IPD manifold and the all-zero vector, and rapidly switch between the manifold (for source frames) and the all-zero vector (for non-source frames), which is possibly not easy. In this work, to facilitate the network learning, we propose to define the target for non-source frames as the mean point of the (complex-valued) DP-IPD manifold, which can be derived as 
\begin{equation}\label{non-source target}
\begin{aligned}
\bar{q}(f)&= \frac{1}{2\pi}\int_{0}^{2\pi} \tilde{B}(f,\theta) d\theta   \\
&=\frac{1}{2\pi}\int_{0}^{2\pi} e^{-j 2\pi v_f d \text{cos}(\theta)/c} d\theta  \\
&= J_0\left(2\pi v_f d / c\right),
\end{aligned}
\end{equation}
where $J_0(\cdot)$ denotes the zero-order Bessel function of the first kind. DP-IPDs are integrated/averaged over all possible $\theta\in[0,2\pi)$ for one microphone pair, which is analogous to computing the spatial coherence of cylindrically isotropic diffuse sound field \cite{Habets2008GeneratingNM}. 
The non-source target values are real numbers as a function of frequency and microphone distance. For one given microphone distance, the non-source target values are computed for the $F$ discrete frequencies using Eq.~(\ref{non-source target}) and then concatenated to form the target vector. 
Fig. \ref{fig:target} shows the non-source target value as a function of frequency for two different microphone distances. 
% To align the dimensions with DP-IPD for network optimization, we sample the spatial coherence across frequencies to form the real part of the target. The imaginary part is then padded with zeros which can be represented as
% \begin{equation}
% \begin{aligned}
% \mathbf{r}(\theta)= [\mathbf{s}(1), \ldots, \mathbf{s}(F), 0, \ldots, 0]^T.
% \end{aligned}
% \end{equation}

% The diffuse sound field is saturated with sound waves traveling in all directions, creating a rich tapestry of time delays, and DP-IPD corresponds to time delay in the space.  The vector $\mathbf{s}$ encompasses propagation information in all directions, mirroring the spatial aggregation of DP-IPD. Consequently, $\mathbf{s}$ is intricately linked with the DP-IPD vector, exhibiting a minimal correlation across all directions with the DP-IPD vector. This distinctive characteristic facilitates the differentiation between silent and non-silent frames. Meanwhile, the connection between the two allows neural networks to achieve rapid transitions from DP-IPD to silent frames, compared to using a vector of zeros. From Equation 6, it can be observed that $S$ is related to the distance between microphones. Examples of two different spacings are provided in Fig. \ref{fig:target}.

To determine whether one source is active or not in one frame, namely conducting the frame-wise per-source activity detection, we compute the following direct-path to noisy magnitude ratio at the reference channel:
\begin{equation}
v_k(n) = \frac{1}{F} \sum_{f=1}^F \frac{|A_r\left(f, \theta_k\right) S_k(n, f)|}{|X_r(n, f)|},
\end{equation}
where $|\cdot|$ denotes absolute value. If $v_k(n)$ is larger than a pre-set threshold, we consider the $k$-th source is active in frame $n$, otherwise inactive. For active source, we use the DP-IPD target vector, otherwise we use the non-source target vector. 


%cascade three full-narrow blocks. The direct-path IPD embedding is output on the narrow-band module, then an average pooling module is used to map STFT frame to DOA frame. Finally, the direct-path IPD embedding is estimated to the real and imaginary parts through a linear layer with an activation function of tanh. 
                                              
%\subsubsection{Full-Narrow Block}
%\label{sec:fn-block}
%One FN block cascades a full-band BLSTM layer and a narrow-band (B)LSTM layer. The frequency BLSTM learns frequency band dependencies as well as full-band spatial cues, and the narrow-band (B)LSTM learns  spatial cues from single frequency band while the time evolution of full-band module output is established. 
%The BLSTM of the full-band module only processed on the frequency axis to ensure that the method proposed in this paper is causal, meanwhile, it’s easy to switch the method to the offline version by changing the LSTM of the narrow-band module to BLSTM.
%We introduce them separately next.

\subsection{Full-Narrow Network Block}
The (two versions of) proposed network consists of a cascade of full-narrow network blocks. One full-band layer plus one narrow-band layer make up the full-narrow block.  
\subsubsection{Full-band BLSTM layer}
%Frequency BLSTM used to learn full-band information and with one Dropout layer to avoid over-fitting. The input of the frequency BLSTM is a sequence along the frequency axis of single time frame, which can  learn the frequency dependence of spatial cues, meanwhile, using BLSTM along the frequency axis can help estimate the IPD for frequency bands with relatively low direct-path energy proportion by smoothing across the frequency bands. Its input $h^f(t)$ can be represented as:

The full-band BLSTM layers process the time frames independently, and all the time frames share the same network weights. The input is a sequence along the frequency axis of one single time frame: 
\begin{equation}
    H^{\text{full}}(n) = (\mathbf{h}(n,1), \dots,\mathbf{h}(n,f), \dots , \mathbf{h}(n,F)),
\end{equation}
where the superscript $^{\text{full}}$ indicates full-band layer, $\mathbf{h}(n,k)\in \mathbb{R}^{D}$ represents the hidden vector for one time-frequency (TF) bin, and $D$ is the hidden dimension. Note that $\mathbf{h}(n,k)$ is the microphone signals for the first full-band layer as shown in Eq.~(\ref{input}), while it is the output of the previous layer for other layers. The full-band layers focus on learning the inter-frequency dependencies of spatial/localization cues. DP-IPD of different frequencies has a strong correlation, as they are all derived from the same TDOA. In addition, the spatial/localization cues of those frequencies with low direct-path energy can be well predicted with the help of other frequencies. The full-band layers do not learn any temporal information, which is left for the following narrow-band layers. 

%spatial cues, meanwhile, using BLSTM along the frequency axis can help estimate the IPD for frequency bands with relatively low direct-path energy proportion by smoothing across the frequency bands.  frequency BLSTM shares parameters across different time frames, which means it is not used to learn dependencies between different time frames. We focus on learning the spatial cues with strong correlation between frequency bands instead.
\subsubsection{Narrow-band (B)LSTM layer}
The narrow-band (B)LSTM layers process the frequencies independently, and all the frequencies share the same network weights. The input is a sequence along the time axis of one single frequency: 
\begin{equation}
    H^{\text{narrow}}(f) = (\mathbf{h}(1,f), \dots,\mathbf{h}(n,f), \dots , \mathbf{h}(N,f)),
\end{equation}
where the superscript $^{\text{narrow}}$ indicates narrow-band layer. The input vector $\mathbf{h}(n,f)$ is the output vector of the previous full-band layer. Estimating the direct-path localization features in narrow-band has been studied in many conventional methods, such as by channel identification in \cite{7533416}, coherence test in \cite{Mohan2008LocalizationOM}, direct-path dominance test in \cite{Nadiri2014LocalizationOM}. The proposed narrow-band layers focus on exploiting these narrow-band inter-channel information. In addition, DP-IPD is time-varying for moving sound source, and the narrow-band layers learn the temporal evolution of DP-IPD as well.
\subsubsection{Skip connections}  
The full-band and narrow-band layers are tailored to emphasize their specific information domains. However, there's a risk of losing narrow-band information after processing through a full-band layer, and vice versa. To circumvent this, skip connections are incorporated to prevent such information loss. As illustrated in Fig.~\ref{fig:method}, this entails concatenating the input sequence of each full-band layer and narrow-band layer with the original input signal's STFT coefficients (after proper dimension transformation).

\subsection{Fixed-Array Model}
For the scenario with a fixed microphone array for training and test, the model architecture is shown in the Fig. \ref{fig:method} (a). Without loss of generality, the reference channel is set as $r=1$. The model takes as input the $M$-channel microphone signals, and  outputs $M-1$ DP-IPD vectors. Specifically, the input is formed by concatenating the real and imaginary parts of multichannel microphone signals as:
\begin{equation}\label{input}
\begin{aligned}
\mathbf{x}[n, f,:]=&\left[ \mathcal{R}\left(X_1(n, f)\right), \mathcal{I}\left(X_1(n, f)\right), \ldots,\right. \\
&\left.\mathcal{R}\left(X_M(n, f)\right), \mathcal{I}\left(X_M(n, f)\right)\right] \in \mathbb{R}^{2 M},
\end{aligned}
\end{equation}
where $[:]$ is an operator to take all values of one dimension of a tensor. 
The input is first processed by full-narrow blocks to extract the spatial feature embedding of multiple sources which can be formulated as:
\begin{equation}
    \mathbf{h}^{\text{fn}} = FNBlocks(\mathbf{x})\in \mathbb{R}^{N\times F \times D}.
\end{equation}
% Considering the correlation between DP-IPD of different sources over a period of time, and convolutional layers focus on local features. 

Then, a convolutional block is used to separate and extract the DP-IPD vector of multiple sources. The structure of the convolutional block is also shown in Fig. \ref{fig:method} (a) which consists of three 2-dimensional causal convolutional layers and two temporal average pooling layers. Convolutional layers are used to capture the local features to separate the microphone pairs and sources. The average pooling layers are used to compress the frame rate.
The activation function of the first two convolutional layers is $relu$, and the activation function of the last convolutional layer is set as $tanh$ for DP-IPD estimation. The final output is obtained as:
\begin{equation}
    \mathbf{Q} = ConvBlock(\mathbf{h}^{\text{fn}})\in \mathbb{R}^{N\times F \times O},
\end{equation}
where $O=2$ (real and imaginary parts) $\times (M-1)$ (microphone pairs) $\times K$ (sources). 

% IPDnet outputs multiple DP-IPDs of multiple sources with the first channel as a reference channel  that the final network output is $\mathbf{h}^o$ transformed in dimension as:
% \begin{equation}
% \begin{aligned}
%        \mathbf{h}^o[n_2,:] = [\mathbf{\hat{r}}_{1,2}(\theta_1),...,\mathbf{\hat{r}}_{1,M}(\theta_1)),...,\\ \mathbf{\hat{r}}_{1,2}(\theta_K),...,\mathbf{\hat{r}}_{1,M}(\theta_K)], 
% \end{aligned}
% \end{equation}
% where $\mathbf{\hat{r}}_{1,m}(\theta_k)$ denote the estimated DP-IPD of $1,m$ microphone pair, the subscripts of $\mathbf{\hat{r}}$ and $\theta$ denote the indices of microphone and sound sources, respectively. 

The network output is defined as a fixed number of $K$ tracks, each track represents one source (can be non-source) and contains $M-1$ estimated DP-IPD vectors. For one time frame, say $n$, the output  can be written as 
\begin{equation}
\begin{aligned}
        \mathbf{Q}[n,:] = [&\mathbf{\hat{q}}_{1,2}(\theta_1),\dots,\mathbf{\hat{q}}_{1,M}(\theta_1)),  \\ &\dots,\mathbf{\hat{q}}_{1,2}(\theta_K),\dots,\mathbf{\hat{q}}_{1,M}(\theta_K)], 
\end{aligned}
\end{equation}
where $\mathbf{\hat{q}}_{1,m}(\theta_k)$ denotes the estimated DP-IPD for the $1,m$ microphone pair and the $k$-th source, at time frame $n$.
Regarding the source permutation problem, the frame-level permutation invariant training (frame-level PIT) is used. The order of microphone pair is fixed according to the microphone index. 

\subsection{Variable-Array Model}

The proposed IPDNet is an end-to-end DP-IPD estimation network that leverages the network to fully learn useful information from the multi-channel microphone signals for removing the interference of such as noise, reverberation and overlapping speech. The network can be easily designed when the microphone array is fixed, namely the input data has a fixed size. However, when processing different arrays using one network with variable input sizes, implementing the end-to-end DP-IPD estimation network is not trivial. This work proposes a pair-wise processing scheme, namely microphone pairs (pair the reference channel with other channels) are individually processed by weight-shared networks, which makes it feasible for processing arbitrary number of microphones (pairs). The intermediate hidden units of all microphone pairs are mean pooled and broadcast to every pair, which makes microphone pairs communicate to each other. Although such microphone pair communication scheme is not as efficient as the one performed in the fixed-array model that directly learns the dependency of multiple microphones, it largely improves the DP-IPD estimation accuracy compared to when there is no communication between microphone pairs.

% Microphone array generalization model is proposed to solve the deep-learning based SSL methods cannot perform well on the array which unseen in the train data. To achieve the cross array generalization, two issues need to be addressed, the first is how neural networks can adapt to different numbers of microphone arrays. A trained network needs to be able to handle microphone signals of any number of channels, which requires the input dimension of the network to dynamically change with the number of microphones. The second is the network's immunity to the microphone permutation, Spatial information should be recorded through the correlation between microphones, rather than the arrangement of a recording microphone array. A novel training scheme is proposed to account for the addressed aspects as the Fig. \ref{fig:method} (b) shown. 

The network architecture of the variable-array model is shown in Fig. \ref{fig:method} (b). Specifically, the reference channel is set as $r=1$, and other microphones are individually paired with the reference channel to form the network input as: 
% For the microphone signal of M-channel, the network input is $\mathbf{x}=[\mathbf{x}_{1,2},\mathbf{x}_{1,3},\ldots,\mathbf{x}_{1,M}]$ which contains $M-1$ microphone pairs. The subscript $_1$ means the first channel is the reference channel. Each microphone
% pair is composed of the reference microphones and one of the other microphones which can be formulated as:
\begin{equation}
\begin{aligned}
\mathbf{x}_{1,m}[n, f,:]=\left[\mathcal{R}\left(X_1(n, f)\right), \mathcal{I}\left(X_1(n, f)\right),\right. \\
\left.\mathcal{R}\left(X_m(n, f)\right), \mathcal{I}\left(X_m(n, f)\right)\right] \in \mathbb{R}^{4},
\end{aligned}
\end{equation}
% This work is applicable to arrays of varying sizes and eliminates the issue of permutations by utilizing pair-wise mapping. 
which is then processed by weight-shared full-narrow blocks:
\begin{equation}
    \mathbf{h}^{\text{fn}}_{1,m} = FNBlocks(\mathbf{x}_{1,m})\in \mathbb{R}^{N\times F \times D}.
\end{equation}
Let $\mathbf{h}^{\text{f/n}}_{1, m} \in \mathbb{R}^{N\times F \times D} $ denotes the hidden units of any one of full-band BLSTM or narrow-band LSTM for the $1,m$ microphone pair.
An average operation after each full-band BLSTM and narrow-band (B)LSTM is conducted as:
\begin{equation}
\mathbf{c}^{\text{f/n}}[n,f,:] =\frac{1}{M-1} \sum_{m=2}^{M} \mathbf{h}^{\text{f/n}}_{1,m}[n,f,:],
\end{equation}
which contains the intermediate information extracted from all microphone pairs.  As the input to the next layer, $\mathbf{c}^{\text{f/n}}$ is first concatenated onto each $\mathbf{h}^{\text{f/n}}_{1, m}$, and then transformed back to $D$-dimensional from $2D$-dimensional via a Linear layer. This interaction between microphone pairs provides the dependency across the entire array, and thus helps each microphone pair learn better. 

Finally, the same convolutional block as in the fixed-array model is used to separate the DP-IPD of multiple sources from the output of full-narrow blocks:
\begin{equation}
    \mathbf{Q}_{1,m} = ConvBlock(\mathbf{h}_{1,m}^{\text{fn}})\in \mathbb{R}^{F\times T \times O'},
\end{equation}
and now $O'=2$ (real and imaginary parts) $\times K$ (sources). 

Similarly to the fixed-array model, for each of the $M-1$ microphone pairs, the network output is defined as a fixed number of $K$ tracks, and each track represents one source (can be non-source). For one time frame $n$, the output can be written as
\begin{equation}
\begin{aligned}
        \mathbf{Q}_{1,m}[n,:] = [\mathbf{\hat{q}}_{1,m}(\theta_1), \dots,\mathbf{\hat{q}}_{1,m}(\theta_K)]. 
\end{aligned}
\end{equation}
The frame-level PIT is also used for training. The order of microphone pair now can be arbitrarily set. 

% Finally, a microphone pair outputs DP-IPD for $K$ tracks, each track represents a source, the final outputs of a microphone pair are as follows:
% \begin{equation}
% \begin{aligned}
%        \mathbf{h}^o_{1,m}[n_2,:] = [\mathbf{\hat{r}}_{1,m}(\theta_1),...,\mathbf{\hat{r}}_{1,m}(\theta_K)], 
% \end{aligned}
% \end{equation}
%  same as one-array-dedicated model, the source permutation problem is solved through frame-level PIT. The outputs of M-channel microphone signal are $[\mathbf{h}^o_{1,2},\mathbf{h}^o_{1,3},\ldots,
% \mathbf{h}^o_{1,M}]$.

\subsection{Frame-level PIT and Sound Source Localization}
The source permutation problem in the training stage is solved through frame-level permutation invariant training (PIT). Let $\alpha\in P$ denote one of the all possible source orders. For one  time frame $n$, the frame-level PIT loss can be computed as:
\begin{equation}
\mathcal{L}(n)= \min _{\alpha \in \operatorname{P}} \frac{1}{K(M-1)} \sum_{k=1}^K \sum_{m=2}^M \operatorname{MSE}\left(\mathbf{q}_{1,m}(\theta_{\alpha(k)}), \hat{\mathbf{q}}_{1,m}(\theta_{{k}})\right)
\end{equation}
% \begin{equation}
% l_{\alpha, n}=\frac{1}{K} \sum_{k=1}^K \operatorname{MSE}\left(\mathbf{q}_{\alpha, s}^*, \hat{\mathbf{q}}_{s}\right),
% \end{equation}
Note that all the $M-1$ microphone pairs share the same source order. The frame-wise PIT loss is averaged over frames as the overall training loss. The mean squared error (MSE) is used as the loss function.

At inference, each track of DP-IPD estimation is associated to one source  (or non-source). Computing the inner product of the estimated DP-IPDs  with the DP-IPD templates, we can derive one source's spatial spectrum: 
\begin{equation}
s_k(i)=\frac{1}{M-1}\sum_{m=2}^M \hat{\mathbf{q}}_{1,m}(\theta_{{k}})^{\top} {\mathbf{q}}_{1,m}(\theta_{i}),
\end{equation}
where $\theta_{i}$ denotes the $i$-th ($i\in[1,I]$) candidate location. The spatial spectrum $s_k(i)$ is independent for each source, as a function of candidate locations.  
If the maximum value of $s_k(i)$ exceeds a pre-defined threshold, a source is deemed present at the corresponding candidate location, otherwise, it is considered as non-source.

\subsection{Configurations}
The proposed network can be easily implemented for both offline or online SSL,  
by setting the narrow-band LSTMs to be bidirectional or unidirectional and the convolutional layers to be non-causal or causal, respectively. 
To make the model easier to optimize, Laplace normalization is performed on the network input as $X_m(n, f) / \mu(n)$, where $\mu(n)$ is a normalization factor. For offline SSL, $\mu(n)$ is computed as $\frac{1}{NF} \sum_{n=1}^N \sum_{f=1}^{F}\left|X_m(n, f)\right|$. For online SSL, $\mu(n)$ is recursively calculated as $\mu(n)=\beta \mu(n-1)+(1-\beta) \frac{1}{F} \sum_{f=1}^{F}\left|X_m(n, f)\right|$ \cite{Yang2022McNetFM} to ensure the causality of the method. Here, $\beta={(L-1)}/{(L+1)}$ denotes the smoothing weight of the historical time frames, and $L$ represents the length of smoothing window.
% \usepackage{multirow}
% \usepackage{multirow}

\section{Experiments and Discussions}\label{sec:exp}
In this section, the performance of the proposed method on both simulated and real-world data are presented. We first describe the experimental setup and then give the experimental results with detailed discussions.

\begin{figure}
    \centering
    \includegraphics[scale=0.58]{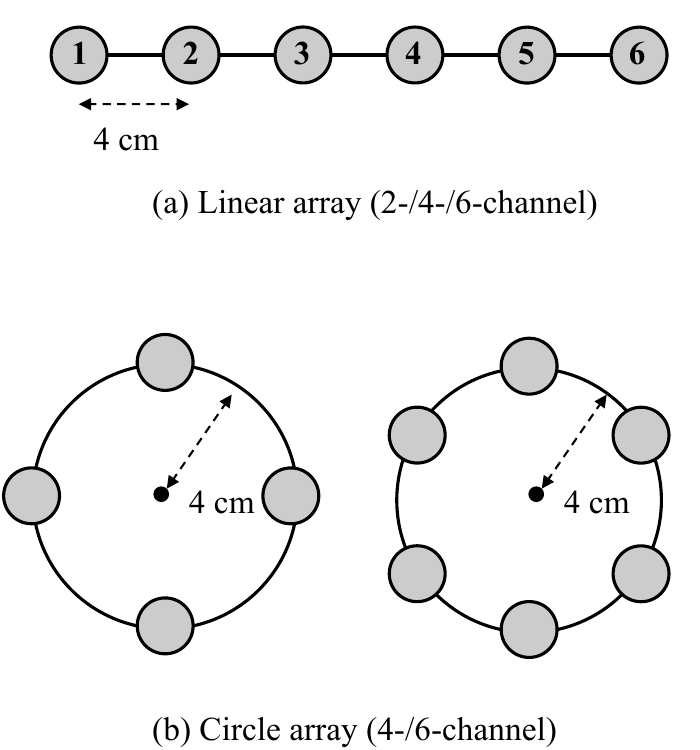}

\caption{Test microphone arrays.}	
\label{fig:mic_array}
\end{figure}

% \begin{figure}
%     \centering
%     \begin{subfigure}{\textwidth} 
%         \includegraphics[scale=0.7]{linear_array.pdf}
%         \caption{Linear array (6-CH)}
%         \label{fig:sub1}
%     \end{subfigure}
    
%     \begin{subfigure}{\textwidth} 
%         \includegraphics[scale=0.65]{circle_array.pdf}
%         \caption{Circle array (4-CH/6-CH)}
%         \label{fig:sub1}
%     \end{subfigure}
%     \caption{Microphone array topologies of the one-array-dedicated data. Note that three linear arrays (2-mic, 4-mic and 6-mic) utilized, with 2-mic (composed of microphones 3 and 5) and 4-mic (composed of microphones 2, 3, 5, and 6) being subarrays of (a).}
%     \captionsetup{justification=centering}
%     \label{fig:mic_array}
% \end{figure}

\subsection{Datasets}
We train the model on simulated datasets, and evaluated it on both simulated and real-world data.

\textbf{Simulated dataset:}  Multichannel microphone signals are simulated through convolving RIRs with speech source signals. RIRs (of moving sources) are generated using the gpuRIR toolbox \cite{DiazGuerra2018gpuRIRAP}. 
Clean speech signals are randomly selected from the training, dev and test sets of the LirbriSpeech corpus \cite{Panayotov2015LibrispeechAA}. Single-source microphone signals are added to obtain multi-source microphone signals. The number of sound sources is set to 2. The room reverberation time (RT60) is randomly set in the range of [0.2, 1.3] s. The room size is randomly set in the range from 6×6×2.5 m to 10×8×6 m. Diffuse noise signals generated following \cite{Habets2008GeneratingNM} are added to  speech signals with a signal-to-noise ratio (SNR) randomly selected from -5 dB to 15 dB. 
For the variable-array model, the microphone arrays used for  training are randomly generated. We pre-defined 6 categories of microphone array, namely uniform/non-uniform linear microphone array, circular microphone array, circular with center microphone array, 2D/3D ad-hoc microphone array. Each category of microphone array is equally presented in proportion. The number of microphones is randomly set in the range of [2, 8]. The distance of any two microphones is limited to [3, 25] cm.  

Five commonly-used microphone array topology are set for test, including 2-/4-/6-channel linear arrays and 4-/6-channel circular arrays, which are shown in Fig. \ref{fig:mic_array}. For the fixed-array model, one network is trained for each test array. The number of utterances (of both the fixed-array and variable-array models) for training, validation and test are 300,000, 4,000, and 4,000, respectively. Each utterance includes two sources, the length of each source's audio clip is in the range of [5, 25] s, and the two audio clips are overlapped according to an overlap rate in the range of [0, 1]. The moving speed of speakers is in the range of [0, 1] m/s.

\textbf{Real-world dataset: } The LOCATA\cite{Lllmann2018TheLC} dataset provides audio signals recorded in the computing laboratory of the Department of Computer Science at the Humboldt University Berlin. The room size is 7.1×9.8×3 m and the reverberation time is 0.55 s. This dataset includes tasks for localizing single/multiple static sound sources (task 1 and 2) and single/multiple moving sound sources (task 3, 4, 5, and 6) using four different microphone arrays. 
% namely benchmark2 (12-mic pseudo-spherical array), DICIT \cite{Brutti2010WOZAD} (15-mic planar array), eigenmike (32-mic spherical microphone array) and the hearing aids (4-mic array). 
We evaluate on all the 6 tasks with the commonly used benchmark2 (12-mic pseudo-spherical array) and DICIT (15-mic planar array) \cite{Brutti2010WOZAD} arrays. Note that we only consider the LOCATA utterances with no more than 2 speakers.

\subsection{Configurations}
The sampling rate of microphone signals is 16 kHz. The window length of STFT is 512 samples (32 ms) with a frame shift of 256 samples (16 ms). The length of training audio clips is 4.5 s. The modeld output a localization result every 12 frames (192 ms). The maximum number of sources is set to $K=2$. The threshold for source activity detection is set to 0.001. The threshold for the  estimated spatial spectrum to determine the presence of a source is set to 0.5.

The number of cascaded Full-Narrow blocks is set to 2. The hidden dimension of the network has been well searched. For the variable-array model, the hidden dimension of every (B)LSTM layers are all set to 128. For the 2-channel and 4-/6-channel fixed-array models, the hidden dimension of every (B)LSTM layers are all set to 128 and 256, respectively, where a higher number of channels carries more information and thus requires a larger network.  
The Adam \cite{Kingma2014AdamAM} is used as the optimizer for training. The batch size of the fixed-array model and variable-array model are set to 16 and 4, respectively. The learning rate is initially set to 0.0005, and exponentially decays with a decaying factor of 0.975. We train the model for almost 30 epochs for the fixed-array model and 15 epochs for the variable-array model.  

\subsection{Evaluation Metrics}
\begin{table}[]

\renewcommand\arraystretch{1.33}
\tabcolsep0.012in
\caption{Results of ablation studies, FB and NB represent full-band and narrow-band, respectively.}

\centering
\label{tab:ablation}
\begin{tabular}{clcccccccccccc}
\toprule \toprule
\multicolumn{2}{c}{}                                            &  &                                                                               &  &                                                                             &  & \multicolumn{3}{c}{Tolerance: 5°}                                                                                                                                       &  & \multicolumn{3}{c}{Tolerance: 10°}                                                                                                                                      \\ \cmidrule{8-14}
\multicolumn{2}{c}{\multirow{-2}{*}{Method}}                    &  & \multirow{-2}{*}{\begin{tabular}[c]{@{}c@{}}\#Params.\\ {[}M{]}\end{tabular}} &  & \multirow{-2}{*}{\begin{tabular}[c]{@{}c@{}}FLOPs\\ {[}G/s{]}\end{tabular}} &  & \begin{tabular}[c]{@{}c@{}}MDR\\ {[}\%{]}\end{tabular} & \begin{tabular}[c]{@{}c@{}}FAR\\ {[}\%{]}\end{tabular} & \begin{tabular}[c]{@{}c@{}}MAE\\ {[}°{]}\end{tabular} &  & \begin{tabular}[c]{@{}c@{}}MDR\\ {[}\%{]}\end{tabular} & \begin{tabular}[c]{@{}c@{}}FAR\\ {[}\%{]}\end{tabular} & \begin{tabular}[c]{@{}c@{}}MAE\\ {[}°{]}\end{tabular} \\ \midrule
\multicolumn{2}{c}{\cellcolor[HTML]{EFEFEF}IPDnet (prop.)} &  & \cellcolor[HTML]{EFEFEF}0.7                                                   &  & \cellcolor[HTML]{EFEFEF}19.4                                               &  & \cellcolor[HTML]{EFEFEF}\textbf{17.1}                  & \cellcolor[HTML]{EFEFEF}\textbf{16.8}                           & \cellcolor[HTML]{EFEFEF}\textbf{1.6}                  &  & \cellcolor[HTML]{EFEFEF}\textbf{8.3}                  & \cellcolor[HTML]{EFEFEF}\textbf{7.7}                   & \cellcolor[HTML]{EFEFEF}\textbf{2.1}                  \\
                        & \qquad w/o FB BLSTM                          &  & 0.5                                                                           &  & 12.7                                                                       &  & 64.4                                                  & 33.5                                                   & 2.1                                                   &  & 51.9                                                   & 21.0                                                   & 3.6                                                   \\
                        & \qquad w/o NB LSTM                           &  & 0.4                                                                           &  & 10.6                                                                       &  & 19.3                                                   & 21.4                                          & 1.7                                          &  & 9.5                                                   & 11.5                                                    & 2.3                                          \\
                        & \qquad w/o Conv. block                       &  & 0.4                                                                           &  & 12.8                                                                       &  & 20.3                                                   & 17.5                                                   & 1.7                                          &  & 10.5                                                   & 7.8                                                    & 2.2 \\ \bottomrule \bottomrule                                         
\end{tabular}
\end{table}
The performance is only evaluated on voice-active periods. The resolution of candidate azimuths and elevations are both 1$\degree$. The angle estimation error is computed as the difference of estimated and true angles. \textit{tolerance: n} means that the source is considered to be successfully localized if the azimuth estimation error is not larger than n$\degree$. Evaluation metrics include miss detection rate (MDR), false alarm rate (FAR) and mean absolute error (MAE). MDR and FAR represent the frame proportions of source active but not successfully localized and source detected but not active, respectively. MAE represents the absolute angle estimation error of all successfully localized sources and time frames.

\subsection{Ablation study}

% \usepackage{multirow}

\begin{figure}
    \centering
    \includegraphics[scale=0.67]{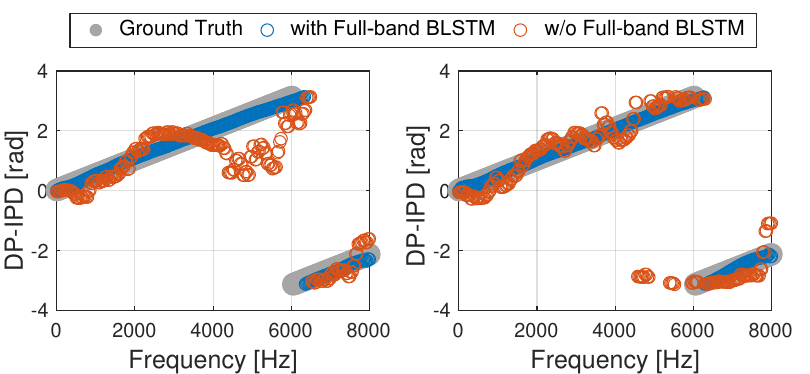}
    \caption{DP-IPD estimations for the proposed model with or without full-band layers. The acoustic condition for left figure is: RT60 = 0.6 s, SNR = 0 dB (white noise) and for right figure is: RT60 = 0.6 s, SNR = 10 dB (white noise).}
    \label{fig:nofull}
\end{figure}
To analyze the contribution of various components of the proposed network, ablation experiments are conducted with the fixed-array model for the simulated 2-channel array, and 180$\degree$-azimuth localization is performed.  Table \ref{tab:ablation} presents the localization performance, model size, and the number of floating point operations (FLOPs) \footnote{The FLOPs, expressed in Giga per second (G/s), are calculated based on 4.5 s long utterances, and then divided by 4.5. For the computation of FLOPs, we utilize the official tool available in PyTorch (torch.utils.flop\_counter.FlopCounterMode on the meta device).} for the proposed network and its variants achieved by removing one sub-network. It is shown that SSL performance noticeably degrades when anyone of the three sub-networks is removed, which indicates the important contribution of the sub-networks to the overall performance. Especially, the full-band BLSTM module seems playing an extremely important role. The full-band BLSTM learns the cross-frequency dependency of localization information. As shown in Eq.~(\ref{dp-ipd}), DP-IPD is basically a linear function of frequency, and leveraging such a strong cross-frequency relationship of DP-IPD would be very helpful for improving the estimation accuracy. Fig. \ref{fig:nofull} shows two examples of DP-IPD estimation. It can be seen that, when SNR is 0 dB, the DP-IPD estimates are highly biased for those frequencies possibly have very low SNR, and the full-band BLSTM helps to correct them based on the cross-frequency relationship of DP-IPD.

% , it can be observed that the method performs poorly, mainly due to the different SNRs of each frequency band. Processing each frequency band independently makes it difficult to estimate an accurate DP-IPD component for bands with low SNR.  DP-IPD of different frequencies has a strong correlation, as they are all derived from the same TDOA. Full-band BLSTM processing on the frequency axis which can learn the the dependence of DP-IPD. We visualized the DP-IPD estimated by the network after removing the full-band BLSTM as the Fig. \ref{fig:nofull} shown, it can be observed that the estimated DP-IPD is chaotic without the full-band BLSTM which consistent with our analysis. Removing the narrow-band LSTM results in notably poorer performance compared to the full IPDnet configuration. This highlights the crucial role of the narrow-band LSTM, which leverages narrow-band inter-channel information and captures the temporal evolution of DP-IPD, in enhancing the system's effectiveness. Meanwhile, Convolutional block is also necessary for IPDnet which is used to capture the local features to separate the source.

\subsection{Comparison with Different Learning Targets}
\label{sec:target}

 \begin{figure}
    \centering
    \includegraphics[scale=0.48]{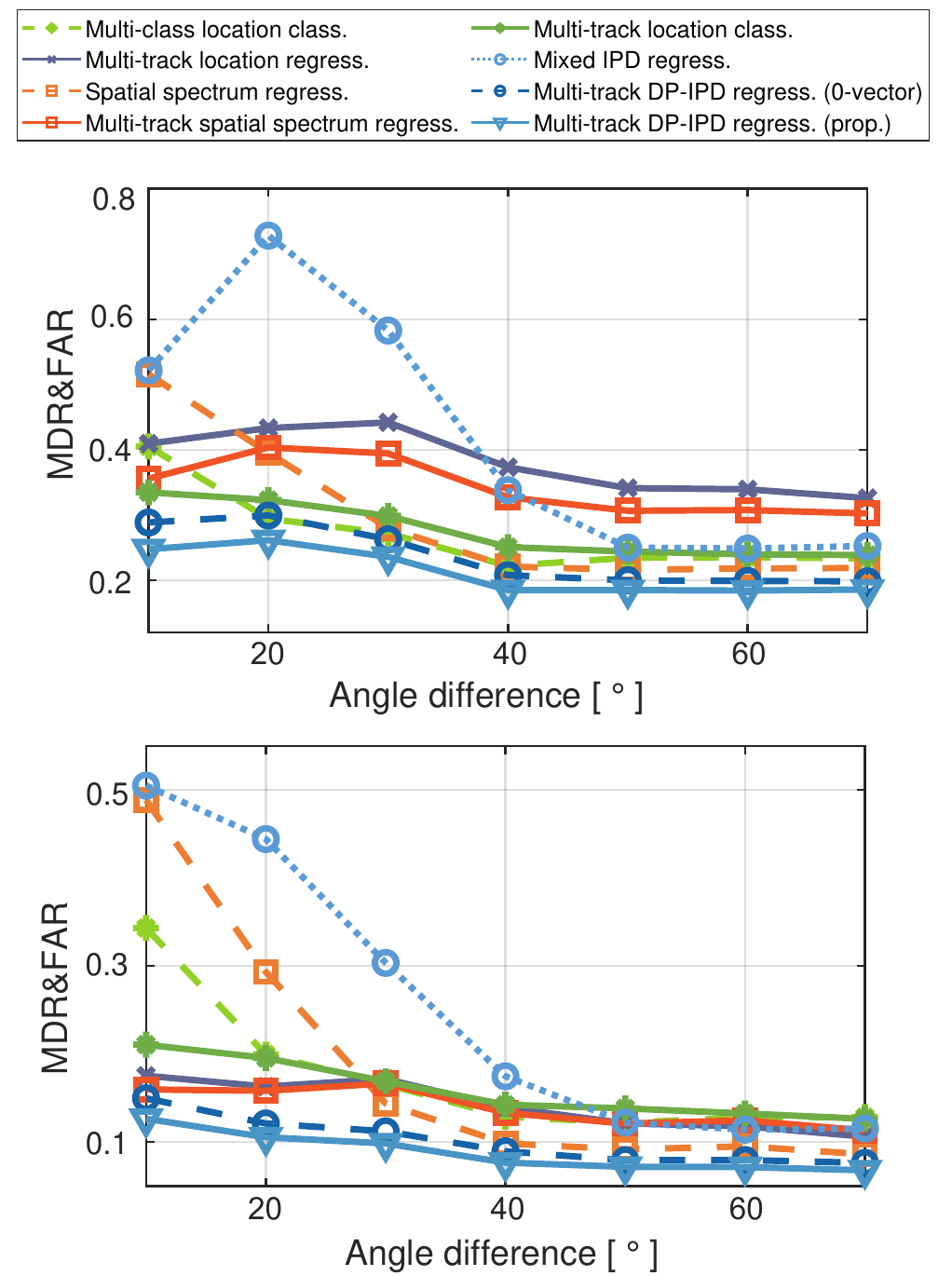}
    \caption{SSL performance for different learning targets, as a function of the angle difference of two sources. The error tolerance is (up) 5\degree, and (bottom) 10\degree.}
    \label{fig:angdiff}
\end{figure}

Deep learning based sound source localization methods use various different learning targets, among which location classification and regression are two commonly used targets. We compare the proposed DP-IPD target with other targets.  Comparison experiments are conducted with the simulated 2-channel array, and 180$\degree$-azimuth localization is performed. The same backbone network as the proposed 2-channel fixed-array model is used, on top of which an extra head will be added if necessary. The following targets are compared, which are all well tuned to achieve their optimal performance on the present task. (1) \textit{Multi-class location classification} \cite{Grumiaux2021SaladnetSM}. One candidate location is considered as one class. Multi-class classification simultaneously localize multiple sources. At inference, classes with output probability larger than a threshold are detected as active sources.  (2) \textit{Multi-track location regression} \cite{Shimada2021MultiACCDOALA} predicts the location of multiple sources. The 2D coordinate on a unit circle is used to represent the azimuth angle, and [0,0] represents non-source. Frame-level PIT is used for training. (3) \textit{Spatial spectrum regression} \cite{He2021NeuralNA} predicts a multi-source spatial spectrum, within which each spectral peak represents one source. At inference, the spectral peaks larger than a threshold are detected as active sources. (4) \textit{Multi-track spatial spectrum regression} \cite{Yin2022MIMODoAnetMI} uses an independent spatial spectrum to represent each source. Two spatial spectrum tracks are arranged in the descending order of the azimuth angles. (5) \textit{Multi-track location classification}. Inspired by the \textit{multi-track spatial spectrum regression} method, we also test one new learning target, i.e. multi-track location classification which outputs two classification tracks, each track represents one source. The two tracks are also arranged in the descending order of the azimuth angles. (6) \textit{Mixed IPD regression} \cite{Yang2022SRPDNNLD} predicts the DP-IPD as in the proposed method, but the DP-IPD of multiple sources are mixed together with a mixing weight (based on the power proportion of each source) for each source. This way circumvents the source permutation problem. At inference, an iterative source detection and localization technique is applied to iteratively extract the DP-IPD of each source. (7) \textit{Multi-track DP-IPD regression (0-vector)}, namely the proposed DP-IPD target, but uses the all-zero vector to represent non-source. (8) The proposed \textit{multi-track DP-IPD regression}. 
For all regression-based methods, the loss function is MSE. Except \textit{Multi-track location regression}, all these targets need to set candidate locations, and they all use every 1$\degree$ as one candidate location.   

\begin{figure}
    \centering
    \includegraphics[scale=0.5]{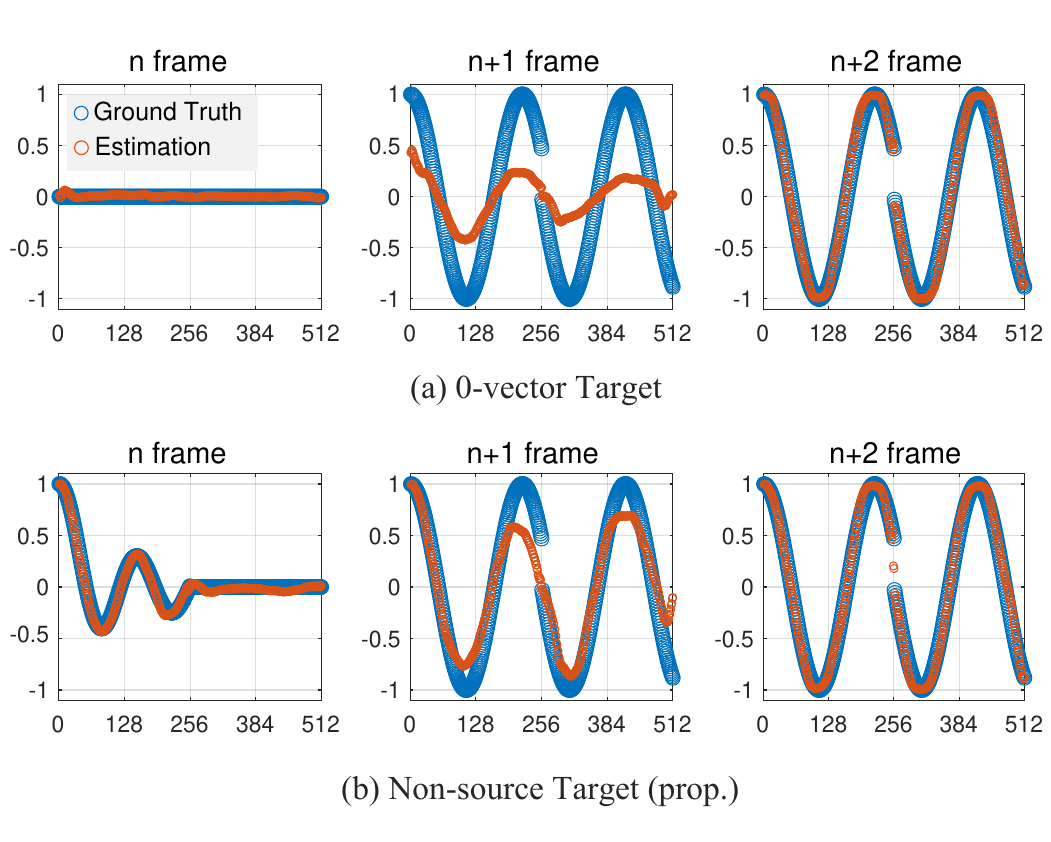}
    \caption{An example of estimated 512-dimensional DP-IPD vector for three consecutive frames where non-source becomes active source at frame $n+1$. }
    \label{fig:newtarget}
\end{figure}

The results (square root of the sum of squared MDR and squared FAR, denoted as MDR\&FAR) as a function of angle difference of the two sources are illustrated in Fig. \ref{fig:angdiff}.
\begin{table*}[]
\renewcommand\arraystretch{1.35}
\scriptsize
\tabcolsep0.032in
\centering
\caption{SSL (azimuth localization) performance on simulated data. Error tolerance is 10$\degree$.}
\label{tab:comexp}
\begin{tabular}{lcllclccccccccccccccccccccc}
\toprule \toprule
\multicolumn{2}{l}{}                                                                                         &                                      &  &                                                                             &  &                                                                               &  & \multicolumn{3}{c}{2-CH}                                                                                                                                                &  & \multicolumn{3}{c}{4-CH LA}                                                                                                                                             &  & \multicolumn{3}{c}{4-CH CA}                                                                                                                                             &  & \multicolumn{3}{c}{6-CH LA}                                                                                                                                             &  & \multicolumn{3}{c}{6-CH CA}                                                                                                                                             \\ \cmidrule{9-27}
\multicolumn{2}{l}{\multirow{-2}{*}{}}                                                                       & \multirow{-2}{*}{Methods}            &  & \multirow{-2}{*}{\begin{tabular}[c]{@{}c@{}}FLOPs\\ {[}G/s{]}\end{tabular}} &  & \multirow{-2}{*}{\begin{tabular}[c]{@{}c@{}}\#Params.\\ {[}M{]}\end{tabular}} &  & \begin{tabular}[c]{@{}c@{}}MDR\\ {[}\%{]}\end{tabular} & \begin{tabular}[c]{@{}c@{}}FAR\\ {[}\%{]}\end{tabular} & \begin{tabular}[c]{@{}c@{}}MAE\\ {[}°{]}\end{tabular} &  & \begin{tabular}[c]{@{}c@{}}MDR\\ {[}\%{]}\end{tabular} & \begin{tabular}[c]{@{}c@{}}FAR\\ {[}\%{]}\end{tabular} & \begin{tabular}[c]{@{}c@{}}MAE\\ {[}°{]}\end{tabular} &  & \begin{tabular}[c]{@{}c@{}}MDR\\ {[}\%{]}\end{tabular} & \begin{tabular}[c]{@{}c@{}}FAR\\ {[}\%{]}\end{tabular} & \begin{tabular}[c]{@{}c@{}}MAE\\ {[}°{]}\end{tabular} &  & \begin{tabular}[c]{@{}c@{}}MDR\\ {[}\%{]}\end{tabular} & \begin{tabular}[c]{@{}c@{}}FAR\\ {[}\%{]}\end{tabular} & \begin{tabular}[c]{@{}c@{}}MAE\\ {[}°{]}\end{tabular} &  & \begin{tabular}[c]{@{}c@{}}MDR\\ {[}\%{]}\end{tabular} & \begin{tabular}[c]{@{}c@{}}FAR\\ {[}\%{]}\end{tabular} & \begin{tabular}[c]{@{}c@{}}MAE\\ {[}°{]}\end{tabular} \\ \midrule
                          &                                                                                  & SRP-DNN\cite{Yang2022SRPDNNLD}                              &  & 2.3                                                                        &  & 0.8                                                                           &  & 19.9                                                   & 13.1                                                   & 2.9                                                   &  & 12.5                                                   & 8.2                                                    & 1.9                                                   &  & 17.1                                                   & 9.0                                                    & 2.5                                                   &  & 12.2                                                   & 6.7                                                    & 1.8                                                   &  & 15.9                                                   & 6.4                                                    & 2.3                                                   \\
                          & \multirow{-2}{*}{\begin{tabular}[c]{@{}c@{}}Fixed-Array\end{tabular}} & \cellcolor[HTML]{EFEFEF}IPDnet (prop.) &  & \cellcolor[HTML]{EFEFEF}19.4/62.8                                       &  & \cellcolor[HTML]{EFEFEF}0.7/2.1                                             &  & \cellcolor[HTML]{EFEFEF}\textbf{8.3}                   & \cellcolor[HTML]{EFEFEF}\textbf{7.7}                   & \cellcolor[HTML]{EFEFEF}\textbf{2.1}                  &  & \cellcolor[HTML]{EFEFEF}\textbf{5.4}                   & \cellcolor[HTML]{EFEFEF}\textbf{3.7}                   & \cellcolor[HTML]{EFEFEF}\textbf{1.4}                  &  & \cellcolor[HTML]{EFEFEF}\textbf{4.7}                   & \cellcolor[HTML]{EFEFEF}\textbf{4.5}                   & \cellcolor[HTML]{EFEFEF}\textbf{1.8}                  &  & \cellcolor[HTML]{EFEFEF}\textbf{4.7}                   & \cellcolor[HTML]{EFEFEF}\textbf{3.5}                   & \cellcolor[HTML]{EFEFEF}\textbf{1.2}                  &  & \cellcolor[HTML]{EFEFEF}\textbf{3.5}                   & \cellcolor[HTML]{EFEFEF}\textbf{4.3}                   & \cellcolor[HTML]{EFEFEF}\textbf{1.6}                  \\ \cmidrule{2-27}
                          &                                                                                  & \cellcolor[HTML]{EFEFEF}IPDnet (prop.) &  & \cellcolor[HTML]{EFEFEF}23.2                                               &  & \cellcolor[HTML]{EFEFEF}1.1                                                   &  & \cellcolor[HTML]{EFEFEF}10.5                           & \cellcolor[HTML]{EFEFEF}12.1                  & \cellcolor[HTML]{EFEFEF}2.3                           &  & \cellcolor[HTML]{EFEFEF}7.6                   & \cellcolor[HTML]{EFEFEF}6.7                   & \cellcolor[HTML]{EFEFEF}1.5                  &  & \cellcolor[HTML]{EFEFEF}5.5                   & \cellcolor[HTML]{EFEFEF}7.8                   & \cellcolor[HTML]{EFEFEF}1.8                  &  & \cellcolor[HTML]{EFEFEF}7.7                   & \cellcolor[HTML]{EFEFEF}5.5                   & \cellcolor[HTML]{EFEFEF}1.3                  &  & \cellcolor[HTML]{EFEFEF}5.1                   & \cellcolor[HTML]{EFEFEF}7.7                   & \cellcolor[HTML]{EFEFEF}1.7                  \\ 
\multirow{-4}{*}{Online}  & \multirow{-2}{*}{Variable-Array}                                                     & \quad w/o mean.                            &  & 21.1                                                                        &  & 0.7                                                                           &  & 9.2                                           & 13.1                                                   & 2.2                                          &  & 8.3                                                    & 9.8                                                    & 1.5                                          &  & 6.6                                                    & 13.0                                                     & 1.9                                                   &  & 8.8                                                    & 8.7                                                    & 1.4                                                   &  & 6.2                                                    & 13.4                                                   & 1.8                                                   \\ \midrule \midrule
                          &                                                                                  & SALADNet\cite{Grumiaux2021SaladnetSM}                             &  & 2.5                                                                        &  & 1.1                                                                           &  &13.5                                                        &10.2                                                        &2.1                                                       &  &9.4                                                        &7.9                                                        &1.5                                                       &  &10.3                                                        &6.0                                                        &1.7                                                       &  &7.3                                                        &7.4                                                        &1.4                                                       &  &9.6                                                        &5.0                                                        &1.5                                                       \\
                          &                                                                                  & SALSA-Lite\cite{Nguyen2021SALSALiteAF}                          &  & 7.5                                                                        &  & 14                                                                            &  & 10.5                                                   & 11.1                                                   & 3.2                                                   &  & 8.4                                                    & 8.4                                                    & 3.0                                                   &  & 10.3                                                   & 7.6                                                    & 3.2                                                   &  & 8.1                                                    & 7.2                                                    & 2.9                                                   &  & 8.2                                                    & 9.1                                                    & 3.1                                                   \\
                          &                                                                                  & SE-Resnet \cite{Han_KU_task3_report}                            &  & 3.3                                                                        &  & 10.2                                                                          &  & 13.6                                                   & 16.5                                                   & 3.3                                                   &  & 10.1                                                   & 14.4                                                   & 2.9                                                   &  & 9.8                                                    & 9.2                                                    & 2.9                                                   &  & 9.0                                                    & 13.6                                                   & 3.0                                                   &  & 10.3                                                   & 8.9                                                    & 3.0                                                   \\
                          & \multirow{-4}{*}{\begin{tabular}[c]{@{}c@{}}Fixed-Array\end{tabular}} & \cellcolor[HTML]{EFEFEF}IPDnet (prop.) &  & \cellcolor[HTML]{EFEFEF}34.6/54.3                                        &  & \cellcolor[HTML]{EFEFEF}0.6/1.8                                             &  & \cellcolor[HTML]{EFEFEF}\textbf{4.6}                   & \cellcolor[HTML]{EFEFEF}\textbf{5.7}                   & \cellcolor[HTML]{EFEFEF}\textbf{1.7}                  &  & \cellcolor[HTML]{EFEFEF}\textbf{3.3}                   & \cellcolor[HTML]{EFEFEF}\textbf{3.6}                   & \cellcolor[HTML]{EFEFEF}\textbf{1.1}                  &  & \cellcolor[HTML]{EFEFEF}\textbf{3.5}                   & \cellcolor[HTML]{EFEFEF}\textbf{3.4}                   & \cellcolor[HTML]{EFEFEF}\textbf{1.3}                  &  & \cellcolor[HTML]{EFEFEF}\textbf{3.2}                   & \cellcolor[HTML]{EFEFEF}\textbf{3.5}                   & \cellcolor[HTML]{EFEFEF}\textbf{0.9}                  &  & \cellcolor[HTML]{EFEFEF}\textbf{2.6}                   & \cellcolor[HTML]{EFEFEF}\textbf{3.9}                   & \cellcolor[HTML]{EFEFEF}\textbf{1.2}                  \\ \cmidrule{2-27}
\multirow{-5}{*}{Offline} & Variable-Array                                                                       & \cellcolor[HTML]{EFEFEF}IPDnet (prop.) &  & \cellcolor[HTML]{EFEFEF}44.5                                               &  & \cellcolor[HTML]{EFEFEF}0.9                                                   &  & \cellcolor[HTML]{EFEFEF}6.9                            & \cellcolor[HTML]{EFEFEF}9.1                            & \cellcolor[HTML]{EFEFEF}1.9                           &  & \cellcolor[HTML]{EFEFEF}4.2                            & \cellcolor[HTML]{EFEFEF}6.9                            & \cellcolor[HTML]{EFEFEF}1.2                           &  & \cellcolor[HTML]{EFEFEF}3.7                            & \cellcolor[HTML]{EFEFEF}6.9                            & \cellcolor[HTML]{EFEFEF}1.4                           &  & \cellcolor[HTML]{EFEFEF}4.3                            & \cellcolor[HTML]{EFEFEF}6.0                            & \cellcolor[HTML]{EFEFEF}1.2                           &  & \cellcolor[HTML]{EFEFEF}3.4                            & \cellcolor[HTML]{EFEFEF}6.8                            & \cellcolor[HTML]{EFEFEF}1.2       \\ \bottomrule \bottomrule                   
\end{tabular}
\end{table*}
% The x-axis of Fig. \ref{fig:angdiff} represents the differences in azimuthal angles between sources, while the y-axis uses the sum of squares of MDR and FAR to indicate the overall performance of localization. A smaller sum of squares for MDR and FAR indicates better localization performance.
% Fig. \ref{fig:angdiff} presents the localization performance of various targets under different angular differences. 
It can be observed that:
\begin{itemize}[leftmargin=*]
\item Multi-track methods outperform their single-track counterparts when the angle difference of two sources is small, for example \textit{multi-track} versus \textit{multi-source spatial spectrum regression}, \textit{multi-track} versus \textit{multi-class location classification}, \textit{multi-track} versus \textit{mixed DP-IPD regression}.  When the two sources are close, the peak of multiple sources in the multi-source spatial spectrum tend to merge into one peak, which however can be well separated by setting two independent tracks. It's worth noting that, the  single-track methods can handle a flexible number of sources, while the multi-track methods can only output a fixed maximum number of sources. 
\item In comparison with classification-based methods, regression-based methods gain a larger accuracy improvement when the error tolerance is increased. For example, comparing \textit{multi-track spatial spectrum regression} with \textit{multi-track location classification}, the latter performs better when the error tolerance is 5$\degree$, while the former performs better when the error tolerance is 10$\degree$. As source location is continuous in space, when the network fails to predict an accurate localization result, using MSE regression loss ensures that the erroneous result does not deviate significantly from the true value. In contrast, the cross-entropy loss lacks such constraint, possibly leading to larger localization error.
\item  The proposed \textit{multi-track DP-IPD regression} consistently outperforms \textit{multi-track location} and \textit{spatial spectrum regression}, while the latter two performs similarly. DP-IPD estimation is a signal-level task, which can be readily learned from  microphone signals. By contrast, location and spatial spectrum estimation require one further step of array-dependent conversion, which may arise more difficulty when directly learned from microphone signals.
\item Compared with using the all-zero vector for non-source, the proposed non-source target achieves better performance, which indicates that it is indeed easier for the network to learn the proposed non-source target.  
% can better learn the representation of silent frames to achieve fast switching from DP-IPD to silent frames. 
To further testify this, in Fig. \ref{fig:newtarget}, we plot an example when switching from one non-source frame to one source frame. 
% The estimated results of three consecutive frames are plotted, where the first frame is a silent frame. 
It can be seen that the proposed non-source target can achieve a quicker transition from the non-source frame to the source frame.
\end{itemize}

\subsection{Results on Simulated Data}

We conduct both online and offline multi-source localization experiments on the simulated dataset.
The following advanced SSL methods are compared with the proposed method: 
\textbf{(1) SRP-DNN \cite{Yang2022SRPDNNLD}} is an online multi-source localization method. It uses a causal CRNN network to estimate the mixed DP-IPD of multiple sources, then the iterative source detection and localization method is used to get the DP-IPD estimate of each source \footnote{https://github.com/BingYang-20/SRP-DNN}. 
\textbf{(2) SALADNet \cite{Grumiaux2021SaladnetSM}} uses cascade convolutional and self-attention modules to perform multi-source  localization. SALADNet originally takes as input feature the intensity vector from the first-order Ambisonics, we change it as the microphone signals in this experiment.
\textbf{(3) SALSA-Lite \cite{Nguyen2021SALSALiteAF}} is designed for joint sound events localization and detection. It uses the frequency normalized IPD concatenated with the magnitude spectrum as the network input, and uses a ResNet-GRU network. We only use its localization branch and change the target from the sound-class-aligned DOA tracks to simply the multi-source DOAs \footnote{https://github.com/thomeou/SALSA-Lite}.
\textbf{(4) SE-Resnet \cite{Han_KU_task3_report}}  is a
top-ranked method for joint sound events localization and detection in DCASE22. It uses a squeeze-and-excitation residual network (as encoder) and a Gated Recurrent Unit network (as decoder). As is done for SALSA-Lite, we change its target to the multi-source DOAs. SALADNet, SALSA-Lite and SE-Resnet perform offline localization.

% Note that, Cross3D and IcoDOA are all designed for online single source localization that we only used them to evaluate the performance on the LOCATA dataset.

% \usepackage{multirow}
% \usepackage{multirow}

Table \ref{tab:comexp} presents the localization performance on five simulated test arrays. The error tolerance is set to 10\degree. Fixed-array models are independently trained for each test array, while the proposed variable-array model is trained once and used for all test arrays.  
% Table \ref{tab:comexp} presents the results of the localization performance. "one-array-dedicated" denotes models trained specifically on a designated microphone array, while "array gen." denotes the microphone array generalization version models. 
In the fixed-array experiments, it can be seen that the proposed method prominently outperforms all the comparison methods under all conditions. The advantages of the proposed method are as follows: i) the proposed method (and SRP-DNN and SALADNet) takes as input the microphone signals, while SALSA-Lite and SE-Resnet take as input the noisy IPD (concatenated with the magnitude spectrum). Directly processing the microphone signals is more effective for suppressing the interference of noise and reverberation, as the natural properties of noise and reverberation (such as their spatial-diffuseness) presented in the original signals can be better leveraged; ii) the proposed learning target, i.e. DP-IPD, is more effective than other learning targets, as discussed in the previous section; iii) the proposed full-band and narrow-band fusion network is efficient to exploit the temporal evolution of narrow-band spatial information and the cross-band correlation of localization cues. 

% In the offline experiment, The proposed method and SALADNet both take as input the microphone signals while SALSA-Lite and SE-Resnet processing the noisy IPD (concatenated with the magnitude spectrum). Directly processing the microphone signals is more effective for suppressing the interferences of noise and reverberation, as the natural properties of noise and reverberation (such as the spatial-diffuseness) presented in the original signals can be better leveraged.  Meanwhile, as the number of microphone increases,
% the localization performance improves, especially when the
% number of microphones increases from 2 to 4. 

We lack direct comparison methods for the proposed variable-array model, but comparing the proposed variable-array model with the fixed-array model still evaluate its efficiency. For the 2-CH array, both the variable-array and fixed-array models directly learn the DP-IPD of one microphone pair within one network. The fixed-array model performs better, since it only handle the fixed 4 cm microphone distance. By contrast, the variable-array model handles a large range of microphone distances, i.e. [3, 25] cm, which requires to cover a more large and difficult learning space. 
For the 4-/6-CH arrays, the pair-wise variable-array model conducts inter-channel communication using the hidden units' mean pooling, which could be sub-optimal compared to the direct inter-channel communication in the fixed-array models. Fortunately, the variable-array model has a reasonable performance degradation relative to the fixed-array models. This indicates that the mean pooling scheme is somehow effective, which can be further testified by the large performance loss when the mean pooling scheme is removed (`w/o mean') as shown in Table \ref{tab:comexp}. In addition, the proposed variable-array model outperforms all other fixed-array comparison methods. This showcases the broader applicability and economic training requirements. Especially, when we consider to use real-recorded data to train the SSL model in the future, there will be no need to collect new data for every new microphone array.  
% we divided the array into microphone pairs to train the microphone array generalized version of the SRP-DNN. It can be seen that our method can adapt to all types microphone array. Although SRP-DNN uses more combinations of microphone pairs, the proposed method outperforms significantly than SRP-DNN on all test sets. 

% In our experiments on microphone array generalization, the term "w/o mean." denotes the omission of the embedding averaging operation for microphone pairs. This is trained separately to highlight the efficacy of our proposed approach in achieving microphone array generalization. It's evident that including the step of concatenating the averaged embeddings significantly aids in the array's generalization capabilities. In comparison with the model dedicated to a single array, both the generalized array model and the "w/o mean." model exhibit a gradual reduction in channel information interaction, which adversely affects localization accuracy. We think that promoting robust interaction among channels is essential for the effective learning of spatial characteristics.

Comparing with other methods, the proposed models have a small model size but a large computational complexity. The proposed full-band/narrow-band network processes frames/frequencies independently, which requires a small model as there is no too much information in one frame/frequency, but the network is run many times and thus has a large complexity. The large computational complexity may limit its use in some real-time applications, and this problem will be resolved in our future work.

% \usepackage{multirow}
\begin{table}[]
\centering
\caption{Azimuth localization performance on all the six tasks of the LOCATA dataset. Error tolerance is 10$\degree$.}
\renewcommand\arraystretch{1.15}
\tabcolsep0.020in
\scriptsize
\begin{tabular}{cclccclccc} \toprule \toprule
                                                                          &                                        &  & \multicolumn{3}{c}{Benchmark2}                                                                                      &  & \multicolumn{3}{c}{DICIT}                                                                                            \\ \cmidrule{4-10} 
\multirow{-2}{*}{}                                                        & \multirow{-2}{*}{Method}               &  & MDR{[}\%{]}                          & FAR{[}\%{]}                          & MAE{[}°{]}                            &  & MDR{[}\%{]}                           & FAR{[}\%{]}                           & MAE{[}°{]}                           \\ \midrule
                                                                          & Cross3D\cite{DiazGuerra2020RobustSS}                                &  & \multicolumn{2}{c}{23.1}                                                    & 3.7                                   &  & \multicolumn{2}{c}{8.8}                                                       & 2.7                                  \\
                                                                          & IcoDOA\cite{DiazGuerra2022DirectionOA}                                 &  & \multicolumn{2}{c}{21.0}                                                    & 3.6                                   &  & \multicolumn{2}{c}{14.7}                                                      & 4.0                                  \\
                                                                          & SRP-DNN\cite{Yang2022SRPDNNLD}                               &  & 0.0                         & 1.5                                  & 1.2                                   &  & 4.2                                   & 4.5                                   & 2.4                         \\
                                                                          & \cellcolor[HTML]{EFEFEF}IPDnet (fixed)&  & \cellcolor[HTML]{EFEFEF}0.0 & \cellcolor[HTML]{EFEFEF}0.0 & \cellcolor[HTML]{EFEFEF}1.5 &  & \cellcolor[HTML]{EFEFEF}1.6           & \cellcolor[HTML]{EFEFEF}1.6           & \cellcolor[HTML]{EFEFEF}2.5          \\
\multirow{-5}{*}{Task 1}                                                  & \cellcolor[HTML]{EFEFEF}IPDnet (variable) &  & \cellcolor[HTML]{EFEFEF}0.0 & \cellcolor[HTML]{EFEFEF}0.0 & \cellcolor[HTML]{EFEFEF}2.5          &  & \cellcolor[HTML]{EFEFEF}0.0  & \cellcolor[HTML]{EFEFEF}0.0  & \cellcolor[HTML]{EFEFEF}2.9          \\ \midrule
                                                                          & SRP-DNN                                &  & 27.8                                 & 3.8                         & 2.4                          &  & 24.4                                  & 10.0                         & 3.0                                  \\
                                                                          & \cellcolor[HTML]{EFEFEF}IPDnet (fixed)&  & \cellcolor[HTML]{EFEFEF}5.5          & \cellcolor[HTML]{EFEFEF}8.0          & \cellcolor[HTML]{EFEFEF}2.7           &  & \cellcolor[HTML]{EFEFEF}{1.1}  & \cellcolor[HTML]{EFEFEF}13.0          & \cellcolor[HTML]{EFEFEF}1.3 \\
\multirow{-3}{*}{Task 2}                                                  & \cellcolor[HTML]{EFEFEF}IPDnet (variable) &  & \cellcolor[HTML]{EFEFEF}4.8 & \cellcolor[HTML]{EFEFEF}8.9          & \cellcolor[HTML]{EFEFEF}4.3           &  & \cellcolor[HTML]{EFEFEF}1.6           & \cellcolor[HTML]{EFEFEF}15.9          & \cellcolor[HTML]{EFEFEF}1.5          \\ \midrule
                                                                          & Cross3D                                &  & \multicolumn{2}{c}{13.9}                                                    & 3.5                                   &  & \multicolumn{2}{c}{15.5}                                                      & 3.2                                  \\
                                                                          & IcoDOA                                 &  & \multicolumn{2}{c}{11.7}                                                    & 3.3                                   &  & \multicolumn{2}{c}{10.6}                                                      & 4.2                                  \\
                                                                          & SRP-DNN                                &  & 1.4                         & 5.8                                  &1.8                          &  & 1.7                                   & 1.7                          & 2.5                                  \\
                                                                          & \cellcolor[HTML]{EFEFEF}IPDnet (fixed)&  & \cellcolor[HTML]{EFEFEF}1.8          & \cellcolor[HTML]{EFEFEF}3.4 & \cellcolor[HTML]{EFEFEF}2.0           &  & \cellcolor[HTML]{EFEFEF}2.6           & \cellcolor[HTML]{EFEFEF}4.8           & \cellcolor[HTML]{EFEFEF}2.0 \\
\multirow{-5}{*}{Task 3}                                                  & \cellcolor[HTML]{EFEFEF}IPDnet (variable)&  & \cellcolor[HTML]{EFEFEF}1.5          & \cellcolor[HTML]{EFEFEF}2.5          & \cellcolor[HTML]{EFEFEF}2.8           &  & \cellcolor[HTML]{EFEFEF}1.2  & \cellcolor[HTML]{EFEFEF}4.2           & \cellcolor[HTML]{EFEFEF}2.1          \\ \midrule
                                                                          & SRP-DNN                                &  & 17.3                                 & 10.0                                 & 2.4                          &  & 17.6                                  & 21.1                                  & 3.4                                  \\
                                                                          & \cellcolor[HTML]{EFEFEF}IPDnet (fixed)&  & \cellcolor[HTML]{EFEFEF}9.2 & \cellcolor[HTML]{EFEFEF}8.5 & \cellcolor[HTML]{EFEFEF}2.4 &  & \cellcolor[HTML]{EFEFEF}7.5  & \cellcolor[HTML]{EFEFEF}14.5 & \cellcolor[HTML]{EFEFEF}2.4 \\
\multirow{-3}{*}{Task 4}                                                  & \cellcolor[HTML]{EFEFEF}IPDnet (variable)&  & \cellcolor[HTML]{EFEFEF}11.3         & \cellcolor[HTML]{EFEFEF}8.9          & \cellcolor[HTML]{EFEFEF}3.4           &  & \cellcolor[HTML]{EFEFEF}8.7           & \cellcolor[HTML]{EFEFEF}16.2          & \cellcolor[HTML]{EFEFEF}2.7          \\ \midrule
                                                                          & Cross3D                                &  & \multicolumn{2}{c}{12.0}                                                    & 3.6                                   &  & \multicolumn{2}{c}{4.5}                                                       & 3.5                                  \\
                                                                          & IcoDOA                                 &  & \multicolumn{2}{c}{11.8}                                                    & 3.6                                   &  & \multicolumn{2}{c}{6.9}                                                       & 3.8                                  \\
                                                                          & SRP-DNN                                &  & 2.3                                  & 15.2                                 & 2.1                                   &  & 0.4                          & 5.3                                   & 2.6                                  \\
                                                                          & \cellcolor[HTML]{EFEFEF}IPDnet (fixed)&  & \cellcolor[HTML]{EFEFEF}1.6 & \cellcolor[HTML]{EFEFEF}2.2          & \cellcolor[HTML]{EFEFEF}2.0  &  & \cellcolor[HTML]{EFEFEF}0.6           & \cellcolor[HTML]{EFEFEF}2.6  & \cellcolor[HTML]{EFEFEF}1.6 \\
\multirow{-5}{*}{Task 5}                                                  & \cellcolor[HTML]{EFEFEF}IPDnet (variable)&  & \cellcolor[HTML]{EFEFEF}3.8          & \cellcolor[HTML]{EFEFEF}1.2 & \cellcolor[HTML]{EFEFEF}3.6           &  & \cellcolor[HTML]{EFEFEF}1.9           & \cellcolor[HTML]{EFEFEF}6.7           & \cellcolor[HTML]{EFEFEF}2.1          \\ \midrule
                                                                          & SRP-DNN                                &  & 8.0                                  & 12.6                                 & 2.7                                   &  & 33.3                                  & 40.6                                  & 3.6                                  \\
                                                                          & \cellcolor[HTML]{EFEFEF}IPDnet (fixed)&  & \cellcolor[HTML]{EFEFEF}7.0 & \cellcolor[HTML]{EFEFEF}5.2 & \cellcolor[HTML]{EFEFEF}2.5  &  & \cellcolor[HTML]{EFEFEF}27.4          & \cellcolor[HTML]{EFEFEF}6.4 & \cellcolor[HTML]{EFEFEF}2.5 \\
\multirow{-3}{*}{Task 6}                                                  & \cellcolor[HTML]{EFEFEF}IPDnet (variable)&  & \cellcolor[HTML]{EFEFEF}11.3         & \cellcolor[HTML]{EFEFEF}5.7          & \cellcolor[HTML]{EFEFEF}3.6           &  & \cellcolor[HTML]{EFEFEF}25.4 & \cellcolor[HTML]{EFEFEF}27.8          & \cellcolor[HTML]{EFEFEF}2.7          \\ \midrule \midrule
                                                                       
                                                                          & Cross3D                                &  & \multicolumn{2}{c}{18.1}                                                    & 3.6                                   &  & \multicolumn{2}{c}{10.1}                                                      & 3.0                                  \\
                                                                          & IcoDOA                                 &  & \multicolumn{2}{c}{16.2}                                                    & 3.5                                   &  & \multicolumn{2}{c}{12.4}                                                      & 3.9                                  \\
                                                                          & SRP-DNN                                &  & \textbf{0.9}                         & 6.0                                    & 1.6                                   &  & 2.9                                   & 3.8                                   & 2.5                                  \\
                                                                          & \cellcolor[HTML]{EFEFEF}IPDnet (fixed) &  & \cellcolor[HTML]{EFEFEF}\textbf{0.9} & \cellcolor[HTML]{EFEFEF}1.4          & \cellcolor[HTML]{EFEFEF}\textbf{1.8}  &  & \cellcolor[HTML]{EFEFEF}1.7           & \cellcolor[HTML]{EFEFEF}2.7           & \cellcolor[HTML]{EFEFEF}\textbf{2.2} \\
\multirow{-5}{*}{\begin{tabular}[c]{@{}c@{}}AVG.\\ (Single \\Source)\end{tabular}} & \cellcolor[HTML]{EFEFEF}IPDnet (variable)&  & \cellcolor[HTML]{EFEFEF}1.3          & \cellcolor[HTML]{EFEFEF}\textbf{0.9} & \cellcolor[HTML]{EFEFEF}2.9           &  & \cellcolor[HTML]{EFEFEF}\textbf{0.6}  & \cellcolor[HTML]{EFEFEF}\textbf{2.2}  & \cellcolor[HTML]{EFEFEF}2.6 
\\  \midrule
   & SRP-DNN                                &  & 7.1                                  & 7.5                                  & 2.0                                   &  & 12.6                                  & 13.4                                  & 2.9                                  \\
                                                                          & \cellcolor[HTML]{EFEFEF}IPDnet (fixed)&  & \cellcolor[HTML]{EFEFEF}\textbf{3.6} & \cellcolor[HTML]{EFEFEF}3.7          & \cellcolor[HTML]{EFEFEF}\textbf{2.1}  &  & \cellcolor[HTML]{EFEFEF}6.9           & \cellcolor[HTML]{EFEFEF}\textbf{9.6}  & \cellcolor[HTML]{EFEFEF}\textbf{2.2} \\
\multirow{-3}{*}{\begin{tabular}[c]{@{}c@{}}AVG.\\ (Multi- \\source)\end{tabular}} & \cellcolor[HTML]{EFEFEF}IPDnet (variable) &  & \cellcolor[HTML]{EFEFEF}4.7          & \cellcolor[HTML]{EFEFEF}\textbf{3.6} & \cellcolor[HTML]{EFEFEF}3.1           &  & \cellcolor[HTML]{EFEFEF}\textbf{6.0}    & \cellcolor[HTML]{EFEFEF}10            & \cellcolor[HTML]{EFEFEF}2.5  

\\ \bottomrule \bottomrule    
\end{tabular}\label{tab:locata}
\end{table}

\subsection{Results on the LOCATA dataset}

\subsubsection{Comparison experiments}
We first evaluate the proposed method and comparison methods for azimuth localization on all the six tasks. We only perform online (causal) SSL according to the setting of LOCATA. Two sub-arrays are used: i) microphone 5, 8, 11, and 12 in the Benchmark2 array, which forms a nearly rectangular array located on the top of robot head; i) microphone 6, 7 and 9 in the DICIT array, which forms a 3-channel linear array with a 4 cm microphone distance. The fixed-array models are re-trained for these two arrays using simulated data. The same variable-array model as used in the previous section is directly used in this experiment. 

For online SSL, besides SRP-DNN \cite{Yang2022SRPDNNLD}, two extra methods are compared: \textbf{(1) Cross3D}  \cite{DiazGuerra2020RobustSS} takes the SRP-PHAT spatial spectrum as input, and uses a causal 3D CNN network to perform moving-source localization \footnote{https://github.com/DavidDiazGuerra/Cross3D}.
\textbf{(2) IcoDOA} \cite{DiazGuerra2022DirectionOA} uses an Icosahedral CNN to extract localization feature from the SRP-PHAT spatial spectrum, and uses a casual CNN to combine the temporal context for moving-source localization \footnote{https://github.com/DavidDiazGuerra/IcoDOA}. Different from the proposed model that automatically detects the number of active speakers, Cross3D and IcoDOA localize fixed one speaker, thus they are compared only on the single-speaker tasks, i.e. task 1, 3 and 5, and uses the localization error rate as evaluation metric. The localization error rate is translated to the equal MDR and FAR.  

 % The LOCATA dataset is used to validate the generalization ability of the proposed method on real-world datasets. We directly apply the trained microphone array generalization version model  to assess performance on the LOCATA dataset, conducting experiments across two different microphone arrays within it. For each microphone array type, we adapt a  (sub-)array configuration: microphones 5, 8, 11, and 12 for Benchmark2, microphones 6, 7, and 9 for DICIT. 

Table \ref{tab:locata} presents the localization performance. 
% All methods were implemented online, and for comparison purposes, each method was trained separately on the dataset corresponding to its respective microphone array. 
% Specifically, IPDnet-O and IPDnet-G denote the model variants tailored for dedicated single-array and microphone array generalization, respectively. 
Across all methods, a consistent bias of approximately 4$\degree$ was observed in the DOA estimation on DICIT data, likely stemming from the annotation bias. To mitigate this effect, we adjusted all DOA estimations by subtracting this bias. It can be seen that the proposed fixed-array model still achieves superior performance compared to other methods, and the proposed variable-array model achieves comparable performance with the proposed fixed-array model. This verifies that the proposed models trained with simulated data can well generalize to real data. Moreover, the proposed variable-array model can well generalize to unseen real microphone arrays.  
% Additionally, it is observed that the microphone array generalization version of the model, trained only once, can perform all tasks on the LOCATA dataset across different microphone configurations. 

% The proposed method and SRP-DNN both take microphone signals as input and predict direct-path localization features, whereas Cross3D and IcoCNN take the noisy SRP-PHAT spatial spectrum as input and predict the source direction. Directly processing the microphone signals proves to be more effective than processing the SRP-PHAT spatial spectrum, particularly in mitigating noise and reverberation interferences.

\subsubsection{Generalization across different number of channels and to elevation estimation} The maximum number of microphones we used for training the variable-array model is 8. In this experiment, we test the variable-array model on a 8-channel and a 12-channel sub-arrays of Benchmark2.  The 8-channel sub-array includes the microphones 1, 3, 4, 5, 8, 10, 11, and 12, which forms a nearly cubic array. The 12-channel array includes all the 12 microphones of Benchmark2, which is a nearly spherical array. In addition, as the 8-channel and 12-channel arrays are both 3D and provide the discrimination ability of vertical direction, the elevation angle is also localized. 
% Candidate locations along elevation are also set as every 1$\degree$, and the DP-IPD templates are theoretically computed using Eq.~(\ref{dp-ipd}).

\begin{figure}[t]
    \centering
    \includegraphics[scale=0.69]{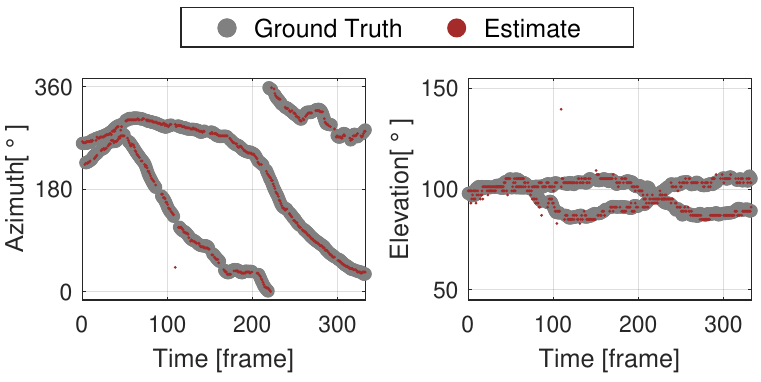}
    \caption{Azimuth and elevation (trajectory) estimations for a LOCATA example with two moving sources. }
    \captionsetup{justification=centering}
    \label{fig:locata}
\end{figure}

Table \ref{tab:number} presents the localization performance, where the results of 4-channel sub-array used in the previous section is also given for comparison. The average performance of all six tasks is reported.  
% the Error tolerance: N$\degree$ indicates that the source is considered successfully localized if the azimuth error is within N$\degree$. 
It can be observed that the performance measures can be gradually improved with the increase of the number of microphones, especially when the error tolerance is 5°, which indicates that more accurate DP-IPD estimation can be obtained with more microphones. This verifies that the proposed variable-array model can well generalize to microphone array with more channels than training arrays. 
The elevation angle is also well localized. This demonstrates that, by separating the feature estimation step and the localization step, the proposed method can be flexibly adapted to various SSL configurations. 
% The metric performance of IPDnet across all sub-arrays is similar. However, when the error tolerance is set to 5°, a noticeable improvement in localization performance can be observed with an increasing number of microphones. 
Fig. \ref{fig:locata} illustrates the localization result of azimuth and elevation for an example with two moving sources, where the 8-channel array is used. It can be seen that both the azimuth and elevation angles can be well localized, but a larger localization error is obtained for elevation.

\begin{table}[t]
\renewcommand\arraystretch{1.25}
\tabcolsep0.05in
\centering
\label{tab:number}
\caption{Azimuth and elevation localization performance using different numbers of microphones of Benchmark2.}
\begin{tabular}{ccccccccccc}
\toprule \toprule
\multirow{2}{*}{Method} &  & \multicolumn{4}{c}{Error tolerance: 5°}                                                                                                                                                                                                     &  & \multicolumn{4}{c}{Error tolerance: 10°}                                                                                                                                                                                                    \\ \cmidrule{3-11}
                        &  & \begin{tabular}[c]{@{}c@{}}MDR\\ {[}\%{]}\end{tabular} & \begin{tabular}[c]{@{}c@{}}FAR\\ {[}\%{]}\end{tabular} & \begin{tabular}[c]{@{}c@{}}AZI\\ {[}°{]}\end{tabular} & \begin{tabular}[c]{@{}c@{}}ELE\\ {[}°{]}\end{tabular} &  & \begin{tabular}[c]{@{}c@{}}MDR\\ {[}\%{]}\end{tabular} & \begin{tabular}[c]{@{}c@{}}FAR\\ {[}\%{]}\end{tabular} & \begin{tabular}[c]{@{}c@{}}AZI\\ {[}°{]}\end{tabular} & \begin{tabular}[c]{@{}c@{}}ELE\\ {[}°{]}\end{tabular} \\ \midrule
IPDnet (4-mic)            &  & 37.1                                                   & 36.1                                                   & 1.9                                                         & -                                                           &  & \textbf{4.7}                                                    & 3.6                                                    & 3.1                                                         & -                                                           \\
IPDnet (8-mic)            &  & 25.1                                                   & 22.7                                                   & \textbf{1.8}                                                        & \textbf{3.4}                                                         &  & 5.5                                                    & \textbf{3.2}                                                    & \textbf{2.5}                                                         & 3.6                                                         \\
IPDnet (12-mic)           &  & \textbf{23.1}                                                   & \textbf{21.6}                                                   & \textbf{1.8}                                                         & \textbf{3.4}                                                         &  & 5.1                                                    & 3.6                                                   & \textbf{2.5}                                                         & \textbf{3.4} \\ \bottomrule \bottomrule

\end{tabular}
\label{tab:number}
\end{table}
% \usepackage{multirow}

\section{Conclusion}\label{conclusion}
This paper proposes a multi-track DP-IPD learning network, named IPDnet, for localization of multiple moving sound sources. The proposed network architecture, i.e. full-band and narrow-band fusion network, is efficient to learn the properties of noise and reverberation and thus to extract reliable DP-IPD of sound sources. The proposed multi-track DP-IPD regression target well disentangles the feature extraction step and the source localization step, and thus outperforms other commonly used SSL targets. Moreover, the proposed variable-array model facilitates the training of SSL network. In this work, the proposed models are trained with pure simulation data in terms of simulated RIR and multi-channel noise, which may has the simulation-to-real problem. To resolve this problem, in future works, the proposed variable-array model can be trained with cross-dataset/array real-recorded data, which provides a reasonable way for alleviating the annotation difficulty and data scarcity when collecting real-world data. 

% IPDnet incorporates a cascaded full-narrow block to effectively fuse both full-band and narrow-band information. The alternating full-band and narrow-band layers are responsible for capturing full-band correlations and extracting narrow-band features of DP-IPDs, respectively. Moreover, the IPDnet for microphone array generalization version is designed which only needs to be trained once and and enabling seamless utilization across different array configurations. This method significantly reduces the burden of data collection and annotation. 
% Experimental results demonstrate the superior performance of IPDnet compared to other advanced methods across simulated and real-world datasets, showcasing its effectiveness in inference with various types of microphone array data.
\bibliographystyle{IEEEtran}
\bibliography{mybib}

\newpage

% your own custom macro containing the $\backslash${\tt{includegraphics}} command to make things
%\bf{If you include a photo:}\vspace{-33pt}
%\begin{IEEEbiography}[{\includegraphics[width=1in,height=1.25in,clip,keepaspectratio]{fig1}}]{Michael Shell}
%Use $\backslash${\tt{begin\{IEEEbiography\}}} and then for the 1st argument use $\backslash${\tt{includegraphics}} to declare and link the author photo.
%Use the author name as the 3rd argument followed by the biography text.
%\end{IEEEbiography}

\vfill

\end{document}